\documentclass[12pt,preprint]{aastex}

\usepackage{natbib}
\usepackage{amsmath}

\newcommand{\arcs}{\mbox{\ensuremath{.\!\!^{s}}}}

\newcommand{\teff}{$T_{{\rm eff}}$}
\newcommand{\vmicro}{\mbox{$\xi_{\rm t}$}}
\newcommand{\gfeh}{\mbox{$[{\rm Fe}/{\rm H}]$}}

\newcommand{\logg}{\mbox{log~{\it g}}}

\begin{document}

\title{The Kepler-454 System: A Small, Not-rocky Inner Planet, a Jovian World, and a Distant Companion}

\author{Sara Gettel\altaffilmark{1}, David Charbonneau\altaffilmark{1}, Courtney D. Dressing\altaffilmark{1}, Lars A. Buchhave\altaffilmark{1,2}, Xavier Dumusque\altaffilmark{1}, Andrew Vanderburg \altaffilmark{1}, Aldo S. Bonomo\altaffilmark{3}, Luca Malavolta\altaffilmark{4,5},  Francesco Pepe\altaffilmark{6}, Andrew Collier Cameron\altaffilmark{7}, David W. Latham\altaffilmark{1},  St\'ephane Udry\altaffilmark{6}, Geoffrey W. Marcy\altaffilmark{8}, Howard Isaacson\altaffilmark{8}, Andrew W. Howard\altaffilmark{9}, 
Guy~R.~Davies\altaffilmark{10,11},
   Victor~Silva Aguirre\altaffilmark{11},
   Hans~Kjeldsen\altaffilmark{11},
   Timothy~R.~Bedding\altaffilmark{12,11}, Eric Lopez \altaffilmark{13}, Laura Affer \altaffilmark{14}, Rosario Cosentino\altaffilmark{15}, Pedro Figueira\altaffilmark{16}, Aldo F. M. Fiorenzano\altaffilmark{15}, Avet Harutyunyan\altaffilmark{15},  John Asher Johnson\altaffilmark{1}, Mercedes Lopez-Morales\altaffilmark{1}, Christophe Lovis\altaffilmark{6},  Michel Mayor\altaffilmark{6}, Giusi Micela\altaffilmark{14}, Emilio Molinari\altaffilmark{15,17}, Fatemeh Motalebi\altaffilmark{6}, David F. Phillips\altaffilmark{1}, Giampaolo Piotto\altaffilmark{4,5}, Didier Queloz\altaffilmark{6,18}, Ken Rice\altaffilmark{13}, Dimitar Sasselov\altaffilmark{1}, Damien S\'egransan\altaffilmark{6}, Alessandro Sozzetti\altaffilmark{3},  Chris Watson\altaffilmark{19},
   Sarbani~Basu\altaffilmark{20},
   Tiago~L.~Campante\altaffilmark{10,11},
   J\o{}rgen~Christensen-Dalsgaard\altaffilmark{11},
   Steven D. Kawaler\altaffilmark{21},
   Travis~S.~Metcalfe\altaffilmark{22},
   Rasmus~Handberg\altaffilmark{11},
   Mikkel~N.~Lund\altaffilmark{11},
   Mia~S.~Lundkvist\altaffilmark{11},
   Daniel~Huber\altaffilmark{12,11},
   William~J.~Chaplin\altaffilmark{10,11}
}

\altaffiltext{1}{Harvard-Smithsonian Center for Astrophysics, 60 Garden Street, Cambridge, Massachusetts 02138, USA; sara.gettel@gmail.com}
\altaffiltext{2}{Centre for Star and Planet Formation, Natural History Museum of Denmark, University of Copenhagen, DK-1350 Copenhagen, Denmark}
\altaffiltext{3}{INAF - Osservatorio Astrofisico di Torino, via Osservatorio 20, 10025 Pino Torinese, Italy}
\altaffiltext{4}{Dipartimento di Fisica e Astronomia ``Galileo Galilei'', Universita'di Padova, Vicolo dell'Osservatorio 3, 35122 Padova, Italy}
\altaffiltext{5}{INAF - Osservatorio Astronomico di Padova, Vicolo dell'Osservatorio 5, 35122 Padova, Italy}
\altaffiltext{6}{Observatoire Astronomique de l'Universit\'e  de Gen\`eve, 51 ch. des Maillettes, 1290 Versoix, Switzerland}
\altaffiltext{7}{SUPA, School of Physics \& Astronomy, University of St. Andrews, North Haugh, St. Andrews Fife, KY16 9SS, UK}
\altaffiltext{8}{University of California, Berkeley, CA, 94720, USA}
\altaffiltext{9}{Institute for Astronomy, University of Hawaii at Manoa, 2680 Woodlawn Dr., Honolulu, Hawaii 96822, USA}
\altaffiltext{10}{School of Physics and Astronomy, University of
  Birmingham, Edgbaston, Birmingham, B15 2TT, UK}

\altaffiltext{11}{Stellar Astrophysics Centre (SAC), Department of
  Physics and Astronomy, Aarhus University, Ny Munkegade 120, DK-8000
  Aarhus C, Denmark}

\altaffiltext{12}{Sydney Institute for Astronomy, School of Physics,
  University of Sydney 2006, Australia}
\altaffiltext{13}{SUPA, Institute for Astronomy, University of Edinburgh, Royal Observatory, Blackford Hill, Edinburgh, EH93HJ, UK}
\altaffiltext{14}{INAF - Osservatorio Astronomico di Palermo, Piazza del Parlamento 1, 90134 Palermo, Italy}
\altaffiltext{15}{INAF - Fundaci\'on Galileo Galilei, Rambla Jos\'e  Ana Fernandez P\'erez 7, 38712 Bre\~na Baja, Spain}
\altaffiltext{16}{Instituto de Astrof\'isica e Ci\^encias do Espa\c co, Universidade do Porto, CAUP, Rua das Estrelas, PT4150-762 Porto, Portugal}
\altaffiltext{17}{INAF - IASF Milano, via Bassini 15, 20133 Milano, Italy}
\altaffiltext{18}{Cavendish Laboratory, J J Thomson Avenue, Cambridge CB3 0HE, UK}

\altaffiltext{19}{Astrophysics Research Centre, School of Mathematics and Physics, Queens University, Belfast, UK}

\altaffiltext{20}{Department of Astronomy, Yale University, New Haven,
  CT, 06520, USA}

\altaffiltext{21}{Department of Physics and Astronomy, Iowa State
  University, Ames, IA 50011, USA}

\altaffiltext{22}{Space Science Institute, 4750 Walnut Street Suite
  205, Boulder, CO 80301, USA}

\slugcomment{Nov. 29, 2015}

\begin{abstract}
Kepler-454 (KOI-273) is a relatively bright ($V$ = 11.69 mag), Sun-like star that hosts a transiting planet candidate in a 10.6 d orbit. From spectroscopy, we estimate the stellar temperature to be 5687 $\pm$ 50 K, its metallicity to be [m/H] = 0.32 $\pm$ 0.08, and the projected rotational velocity to be $v$ sin $i$ $<$2.4 km s$^{-1}$. We combine these values with a study of the asteroseismic frequencies from short cadence $Kepler$ data to estimate the stellar mass to be 1.028$^{+0.04}_{-0.03}$ $M_{\odot}$, the radius to be 1.066 $\pm$ 0.012 $R_{\odot}$ and the age to be 5.25$^{+1.41}_{-1.39}$ Gyr.
 We estimate the radius of the 10.6 d planet as 2.37 $\pm$ 0.13 $R_{\oplus}$. Using 63 radial velocity observations obtained with the HARPS-N spectrograph on the Telescopio Nazionale Galileo and 36 observations made with the HIRES spectrograph at Keck Observatory, we measure the mass of this planet to be 6.8 $\pm$ 1.4 $M_{\oplus}$. We also detect two additional non-transiting companions, a planet with a minimum mass of 4.46 $\pm$ 0.12 $M_{J}$ in a nearly circular 524 d orbit and a massive companion with a period $>$10 years and mass $>$12.1 $M_{J}$. \text{The twelve exoplanets with radii $<$2.7 $R_{\oplus}$} and precise mass measurements appear to fall into two populations, with those $<$1.6 $R_{\oplus}$ following an Earth-like composition curve and larger planets requiring a significant fraction of volatiles. With a density of 2.76 $\pm$ 0.73 g cm$^{-3}$, Kepler-454b lies near the mass transition between these two populations and requires the presence of volatiles and/or H/He gas. 

\end{abstract}

\section{Introduction}
The NASA $Kepler$ mission has detected thousands of planet candidates with radii between 1 and 2.7 $R_{\oplus}$ \citep{bor11,bat13,bur14,row15}. The corresponding population of low mass planets was previously detected in radial velocity surveys \citep{may08,how09}, but notably has no analog in our own solar system.
The composition of these objects is not yet well understood; theoretical models predict that some of these intermediate size planets may be predominantly rocky and others may have a large fractional composition of volatiles or a substantial hydrogen envelope \citep{leg04,val06,sea07,for07,zen13}. There is presently only a small number of such planets observed to transit and having published mass estimates with a precision better than 20\%, while less precise mass estimates are generally not sufficient to distinguish between a rocky compositional model and one that is volatile-rich.

\citet{dre15b} raised the intriguing possibility that the small planets with well-measured masses of $<$6 $M_{\oplus}$, equivalent to about 1.6 $R_{\oplus}$, have similar compositions, well approximated by a two-component model with the same MgSiO$_{3}$/Fe ratio as the Earth \citep{zen13}. Planets larger than 2.0 $R_{\oplus}$ are observed to have lower densities, consistent with a significant fraction of volatiles or H/He gas and do not follow a single mass-radius relation. Due to the limited number of planets smaller than 2.7 $R_{\oplus}$ with precise masses, it is not yet clear how broadly applicable the iron-magnesium silicate model might be.

KOI-273 (KIC 3102384) is a moderately bright solar-like star, with $V = 11.69$ mag and $K_{p} = 11.46$ mag. It was observed by $Kepler$ during Quarters 0-17, with short cadence data taken during Quarters 4, 6-12 and 15-17. It was identified by the $Kepler$ pipeline as having a planet candidate in the first four months of long cadence data, with radius $R_{p}$ = 1.86 $R_{\oplus}$ and period $P$ = 10.57 d. 
This initial radius measurement of KOI-273.01 falls in between the two classes of planets discussed in \citet{dre15b} and so would provide a test case of the Earth-composition mass-radius relation.  

The stellar parameters of KOI-273 were previously determined by a combination of asteroseismology and spectroscopy in \citet{hub13}. They reported an asteroseismically determined stellar radius $R_{*}$ = 1.081 $\pm$ 0.019 $R_{\odot}$, stellar mass $M_{*}$ = 1.069 $\pm$ 0.048 $M_{\odot}$ and surface gravity log $g$ = 4.399 $\pm$ 0.012. Additionally, \citet{hub13} use Stellar Parameter Classification \citep[SPC]{buc12,buc14} on spectra from Keck-HIRES \citep{vog94}, the TRES spectrograph on the 1.5m at Whipple Observatory \citep{fur08} and the Tull spectrograph on the 2.7m at McDonald Observatory \citep{tul95}, to obtain a metallicity of [Fe/H] = 0.350 $\pm$ 0.101 and effective temperature \teff = 5739 $\pm$ 75 K. \citet{mcq13b} searched for the stellar rotation period of KOI-273 in the autocorrelation function of the Kepler photometry and were unable to detect it.

In this paper we measure the mass of KOI-273.01, determine a spectroscopic orbit for an additional Jovian planet and constrain the orbit of a widely-separated companion, by analyzing the 2014 \& 2015 seasons of HARPS-N radial velocities and several seasons of HIRES radial velocities. As this work confirms the planetary nature of KOI-273.01, we adopt the convention of referring to it as Kepler-454b, and the two more distant companions as Kepler-454c and Kepler-454d, respectively. In Sections 2 \& 3, we analyze the spectroscopic and asteroseismic parameters of the star, respectively. In Section 4 we model the Kepler transit photometry. In Section 5, we discuss our radial velocity observations and the data reduction process.  In Section 6,  we develop a radial velocity model to describe the Kepler-454 system. In Section 7, we conclude with a discussion of the mass measurement of Kepler-454b in the context of the bulk densities of small planets.

\section{Spectroscopic Analysis of Stellar Parameters}

We used SPC to derive the stellar parameters of the host star from high-resolution, high signal to noise ratio (SNR) HARPS-N spectra, with an average SNR per resolution element of 85. More details on these observations are provided in Section 5. We ran SPC both with all parameters unconstrained and with the surface gravity constrained to the value determined by asteroseismology \citep[log $g$ = 4.40 $\pm$ 0.01]{hub13}. The surface gravity from the unconstrained SPC analysis, log $g$ = 4.37 $\pm$ 0.10,  is in close agreement with the asteroseismic value. For the final parameters, we constrained the surface gravity to the value from asteroseismology. The weighted mean of the SPC results from the individual spectra yielded \teff = 5687 $\pm$ 50 K, [m/H] = 0.32 $\pm$ 0.08 and $v$ sin $i$ $<$2.4 km s$^{-1}$. 

We also determined the atmospheric parameters using the line analysis code MOOG (Sneden 1973, version 2014) and a Kurucz model atmosphere with the new opacity distribution function (ODFNEW; Castelli \& Kurucz 2004; Kurucz 1992), as done in Dumusque et al. (2014). We measured the equivalent widths of iron spectral lines on a coadded spectrum of SNR $\simeq$ 400 using ARES v2 with automatic continuum determination (Sousa et al. 2015). We used the linelist from Sousa et al. (2011), with the log $gf$ modified to account for the solar iron abundance adopted in MOOG (log~{\it $\epsilon$}(Fe) = 7.50). We obtained \teff = 5701 $\pm$ 34 K, surface gravity $\logg$ = 4.37 $\pm$ 0.06 , microturbulent velocity $\vmicro$ = 0.98 $\pm$ 0.07 km s$^{-1}$, and iron abundance $\gfeh$ = +0.27 $\pm$ 0.04, where the errors include the dependence of the parameters on temperature. Constraining the gravity to the asteroseismology value did not produce any change in the other parameters. Both sets of stellar parameters are summarized in Table \ref{table:stellar}.

We used the relationship in \citet{mam08} to estimate the rotational period of Kepler-454 from log $R'_{\rm HK}$ and $B-V$ = 0.81, resulting in a value of 44.0 $ \pm$ 4.4 d. This estimate is consistent with the minimum rotation period of 23 d we would obtain by combining the stellar radius and upper limit on the projected rotational velocity from SPC, if we assumed that the rotational axis is aligned with the orbital plane of the inner planet.

\section{Asteroseismic estimation of fundamental stellar properties}
 \label{sec:prop}

 \subsection{Estimation of individual oscillation frequencies}
 \label{sec:freq}

The detection of solar-like oscillations was first reported in Kepler-454
by \citet{hub13}. That study used just one asteroseismic
measured parameter -- the average large frequency separation,
$\Delta\nu$, between overtones -- to model the star. Here, we perform
a more detailed analysis, using frequencies of 12 individual modes
spanning seven radial overtones.

The results come from the analysis of \emph{Kepler} short-cadence
data, which are needed to detect the short-period oscillations shown
by the star. Kepler-454 was observed in short-cadence in \emph{Kepler}
observing quarters 4, 6 through 12, and 15 through 17.  A lightcurve
was prepared for asteroseismic analysis using the KASOC filter
\citep{han14}. This mitigates the planetary transits, and
minimizes the impact of instrumental artifacts and noise.

\begin{figure*}
\epsscale{1.0}
\plottwo{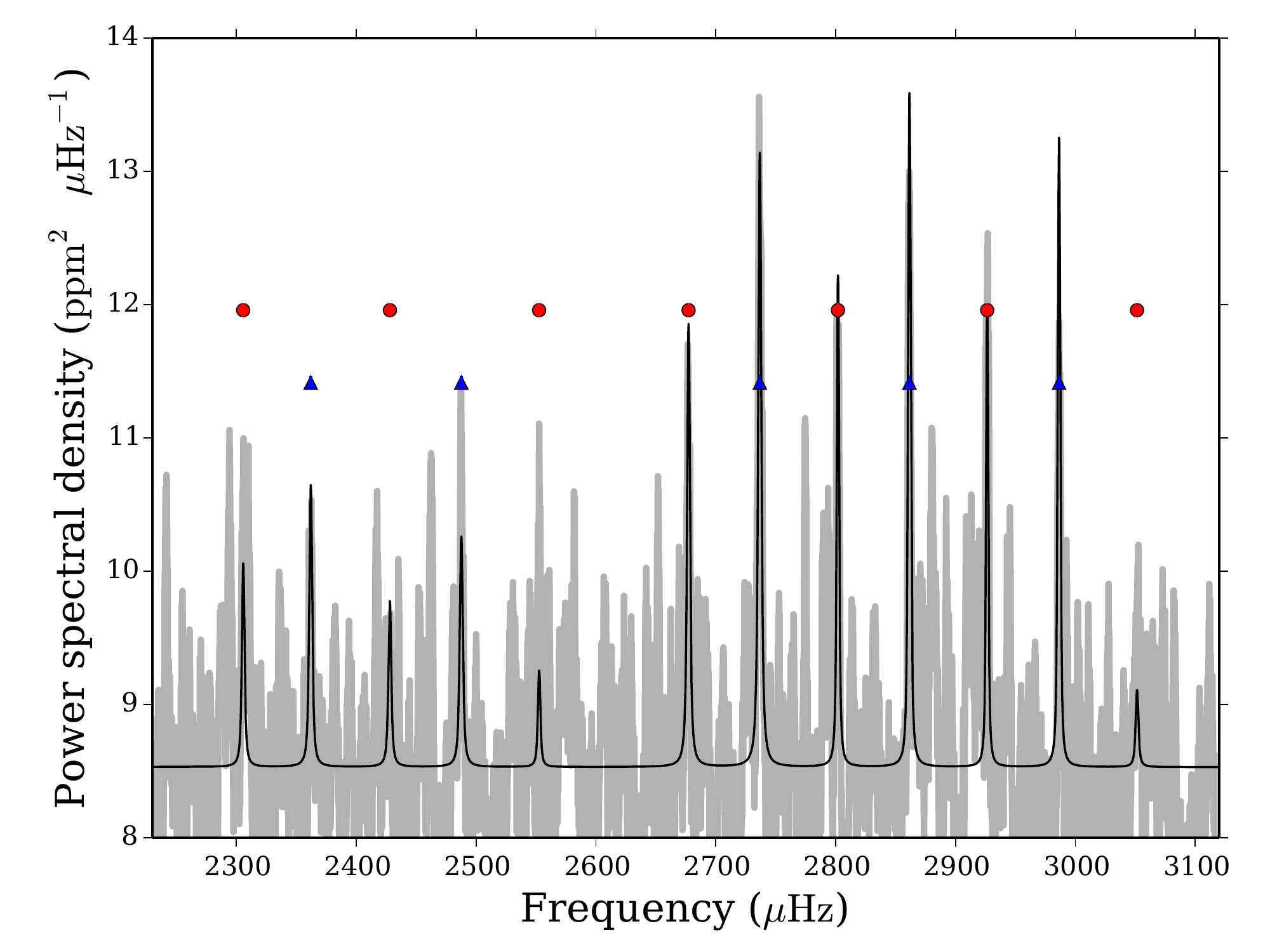}{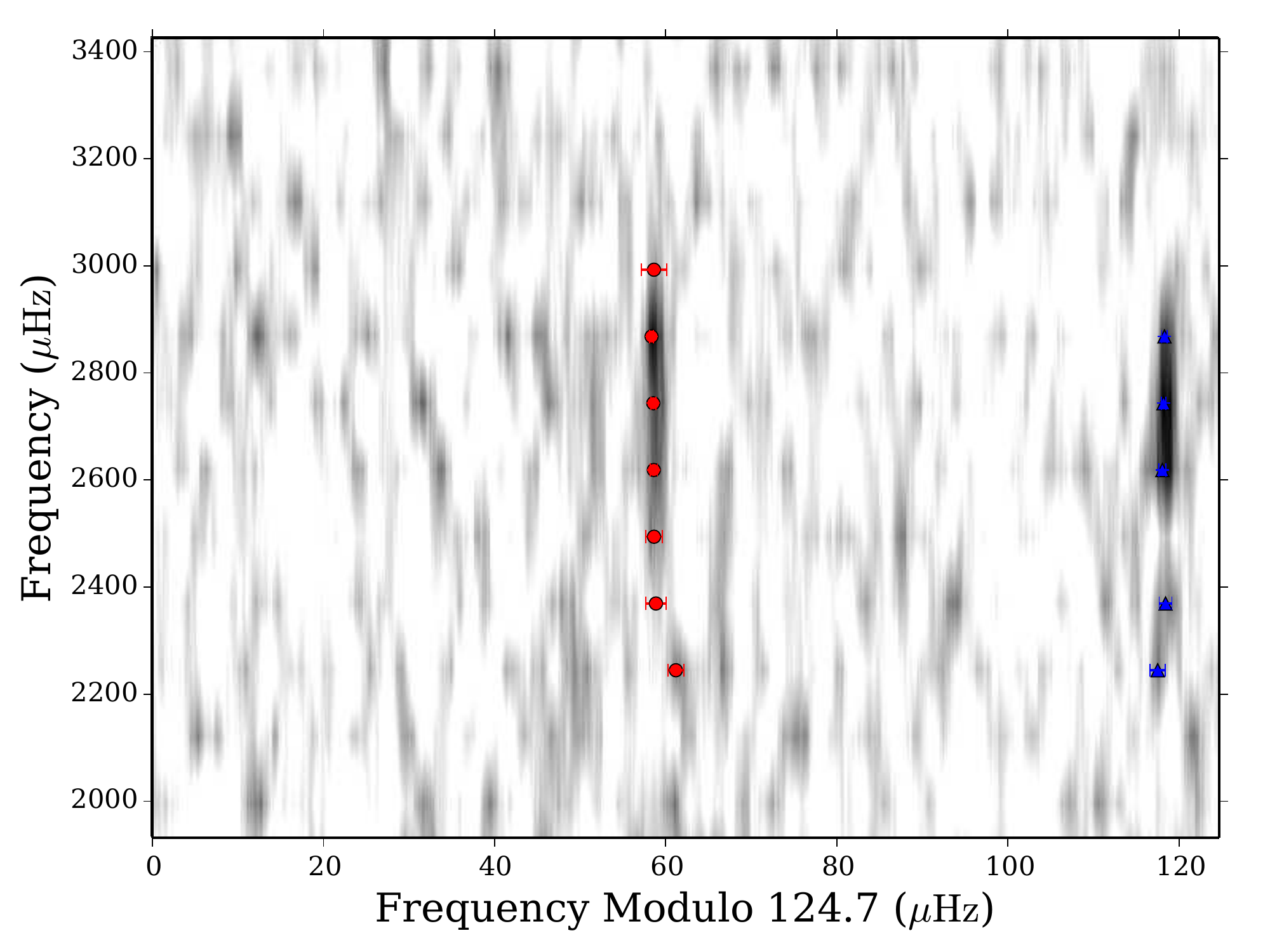}

\caption{\textbf{Left:} Power spectrum of Kepler-454 (gray), after smoothing with
  a boxcar filter of width $3\,\rm \mu Hz$. The best-fitting model of
  the oscillation spectrum is plotted in black. Symbols mark the
  best-fitting frequencies of the $l=0$ modes (red circles), and $l=1$
  modes (blue triangles). \textbf{Right:} \'Echelle diagram of the oscillation
  spectrum of Kepler-454. The spectrum was smoothed with a 3-$\rm \mu Hz$
  Gaussian filter. The best-fitting frequencies are overplotted, with
  their associated $1\sigma$ uncertainties.}

\label{fig:fig1}
\end{figure*}

The left-hand panel of Fig.~\ref{fig:fig1} shows the power spectrum of
the prepared lightcurve after smoothing with a 3-$\rm \mu Hz$ wide
boxcar filter. The plotted range shows peaks due to acoustic
(pressure, or p) modes of high radial order.  The right-hand panel of
Fig.~\ref{fig:fig1} shows the \'echelle diagram, made by dividing the
power spectrum into frequency segments of length equal to
$\Delta\nu$. When arranged vertically, in order of ascending
frequency, the diagram shows clear ridges, comprising overtones of
each angular degree, $l$.

Despite the low S/N, it was possible to extract estimates of
individual frequencies.  Mode identification was performed by noting
that the ridge in the centre of the \'echelle diagram invariably
corresponds to $l=0$ in stars with such values of $\Delta\nu$ \citep{whi11,whi12}.

We produced a set of initial frequencies using a Matched Filter
Response fit to an asymptotic relation for the p-mode spectrum, which
located frequencies along the ridges in the \'echelle diagram
\citep{gil11}.  We then used the ``peak bagging''
methodology described by Davies et al. (2015) to extract the
individual frequencies. We performed a Markov Chain Monte Carlo (MCMC)
optimization fit of the power spectrum, with the
oscillation spectrum modeled as a sum of Lorentzian profiles. The
adopted procedures have recently been used to fit 33 \emph{Kepler}
planet-hosting stars, and full details are given in \citet{dav15}.

The estimated frequencies are plotted in both panels of
Fig.~\ref{fig:fig1}, with the $l=0$ modes shown as red circles, and
$l=1$ modes as blue triangles. Some weak power is present in the $l=2$
ridge, but estimates of the $l=2$ frequencies are marginal and we have
elected not to use them in our modeling.  Best-fitting frequencies and
equivalent $1\sigma$ uncertainties are given in Table~\ref{tab:tab1}.
The reported frequencies have been corrected for the Doppler shift
caused by the motion of the star relative to the observer (here, a
shift of $-71\,\rm km\,s^{-1}$), using the prescription in \citet{dav14}. The correction for the most prominent modes is about
$0.6\,\rm \mu Hz$.

Finally, it is worth noting that the best-fitting frequencies match
very closely the observed low-$l$ solar p-mode frequencies after the
latter have been scaled homologously \citep{bed10} by the
ratio of the respective average large separations. This is not
surprising given the similarity in mass and age to the Sun, and also
confirms that we have correctly assigned the $l$ values.

\subsection{Detailed modeling of the host star}
\label{sec:model}

Stellar properties were determined by fitting the spectroscopic
constraints and extracted oscillation frequencies to several sets of
models using different techniques. These techniques are based on the
use of the individual frequencies or combinations of frequencies to
determine the best-fitting model and statistical uncertainties. Here,
we used four different techniques, in the same configurations that
were employed to model 33 \emph{Kepler} exoplanet host stars \citep{sil15}: the BAyesian STellar Algorithm (BASTA), coupled
to grids of GARSTEC \citep{wei08} stellar evolutionary
models; the ASTEC Fitting (ASTFIT) method and the Asteroseismic
Modeling Portal (AMP), both coupled to ASTEC \citep{chr08} models; and the Yale Monte-Carlo Method (YMCM), coupled to YREC
\citep{dem08} models. Detailed descriptions of the techniques
were given by \citep{sil15}.

We found excellent agreement, at the level of precision of the data,
in the stellar properties estimated by the different techniques. The
final properties listed in Table~\ref{table:stellar} are those obtained
from combining grids of GARSTEC models with the BASTA code, and
include the effects of microscopic diffusion and settling. The quoted
uncertainties are the formal, statistical uncertainties.

We additionally consider the combined effect of different systematic uncertainties, namely the use of different sets of asteroseismic observables, different evolutionary and pulsation codes, as well as different choices of input physics. The expected magnitude of these systematic uncertainties (again, see Silva Aguirre et al. 2015) is as follows: 0.3\% (density and radius), 1\% (mass), and 7\% (age) due to the choice of asteroseismic observables (individual frequencies or combination of frequencies); 1\% (density and radius), 2\% (mass), and 9\% (age) due to the choice of technique; 0.8\% (density), 0.7\% (radius), 2.3\% (mass), 9.6\% (age) due to the choice of input physics, and 1.7\% (density), 1.6\% (radius), 3.6\% (mass), and 16.8\% (age) due to the choice of initial helium abundance. The formal uncertainties in Table 1 are derived in the manner of \citet{sil15} and do not include the systematic effects discussed above.

\section{Photometry}

Our photometric analysis of Kepler-454 is based on 30 months of short cadence data acquired between quarters Q4-Q17 (excluding Q5, Q13, and Q14) and includes 76~independent transits. 
As in \citet{dre15}, we normalized each transit to remove the effects of long term drifts by fitting a linear trend to the out-of-transit light curve surrounding each transit. Specifically, we used the time intervals $1-3.5$ transit durations prior to and following the expected transit center. We divided the flux data by this trend to produce a normalized light curve. 

We used a transit model based on \citet{man02} and varied the period $P$, the epoch of transit $T_{\rm 0}$, the ratio of the semi-major axis to stellar radius $a/R_\star$, the ratio of planet to stellar radius $R_p/R_\star$, and the impact parameter $b = a/R_\star \cos i$, where $i$ is the orbital inclination. We assumed a circular orbit model and fit for quadratic limb darkening coefficients using the parameterization suggested by \citet{kipping2013} in which the parameters $q_1 = (u_1 + u_2)^2$ and $q_2 = 0.5 u_1/(u_1+u_2)$ are allowed to vary between 0 and 1.

We constrained the transit parameters by performing a MCMC analysis with a Metropolis-Hastings acceptance criterion \citep{met53}, starting with the initial parameters provided by \citet{bat13}. We initialized the chains with starting positions set by perturbing the solution of the preliminary fit by up to $5\sigma$ for each parameter. The step sizes for each parameter were adjusted to achieve an acceptance fraction between 10 and 30\%. 

We ran each chain for a minimum of $10^4$ steps and terminated them once each parameter had obtained a Gelman-Rubin reduction factor $\hat{R} < 1.03$. This statistic compares the variance of a parameter in an individual Markov chain to the variance of the mean of that parameter between different chains \citep{gel92,for05} and is used to identify chains that have not yet converged. A value of \textit{\^R} $>$ 1.1 suggests that a chain has not converged \citep{gil95}. While lower values of \textit{\^R} are not a definite indicator of convergence, we select a stopping criteria of \textit{\^R} $<$ 1.03 as a balance between probability of convergence and computational efficiency. 

\begin{figure*}
\centering
\includegraphics[width=0.9\textwidth]{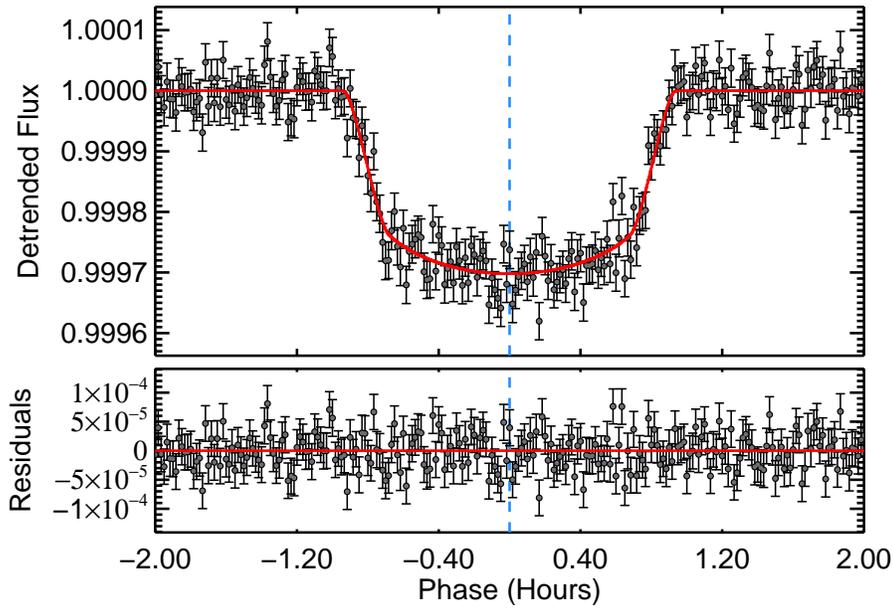}
\caption{\textbf{Top:} Detrended, normalized, phased short cadence observations of Kepler-454 from transits binned to one-minute intervals (data points with errors). The adopted transit model is shown in red and the transit center is marked by the dashed blue line. \textbf{Bottom: } Residuals to the transit fit. \label{lcphase}}
\end{figure*}

We removed all steps prior to the step at which the likelihood first exceeded the median likelihood of the chain, to account for burn-in. We merged the chains and used the median values of each parameter as the best-fit value. We chose the error bars to be symmetric and span the 68\% of values closest to the best-fit value. These values are shown in Table \ref{table:tparam} and the fit to the short-cadence observations is shown in Figure \ref{lcphase}.

Our best-fit orbital period and time of transit agree well with those reported in the Cumulative Kepler Object of Interest table on the NASA Exoplanet Archive\footnote{\url{http://exoplanetarchive.ipac.caltech.edu/cgi-bin/TblView/nph-tblView?app=ExoTbls\&config=cumulative}} as of 19 May 2015, but we found a significantly larger value for the impact parameter. In contrast to the previously reported value of $b = 0.1^{+0.3703}_{-0.0999}$, we find $b = 0.929\pm0.009$, which is $2.2\sigma$ discrepant from the previous value. Our results for $a/R_\star$ and $R_p/R_\star$ are also in disagreement with the KOI table results: we estimate $a/R_\star = 18.293\pm1.098$ and $R_p/R_\star = 0.0204\pm0.001$ whereas the cumulative KOI list reports $a/R_\star = 48.17\pm3.6$ and $R_p/R_\star = 0.015743^{+0.0005}_{-0.000202}$. 

The larger value for $R_p/R_\star$ translates directly into an inferred planetary radius that is significantly larger than the $R_p = 1.86\pm0.03$ $R_{\oplus}$ value reported in the cumulative KOI table. Combining the uncertainty in $R_p/R_\star$ and the uncertainty on the stellar radius ($1.06568\pm 0.01200$ $R_{\odot}$), we find a planet radius of $2.37\pm0.13$ $R_{\oplus}$. Given that our values are consistent with those found by \citet{sliski+kipping2014} in an independent analysis of the Kepler-454b short cadence data, we attribute the disagreement between the results of our analysis and the values reported in the Cumulative KOI table to the degraded ability of analyses based on long cadence data to tightly constrain the impact parameter. In addition, the use of fixed limb darkening parameters for the fit reported in the Cumulative KOI Table may have contributed to the discrepancy.

In addition to revising the transit photometry for Kepler-454b, we searched for transits due to additional companions using a Box-fitting Least Squares analysis \citep{kov02}, but none were found. We also investigated the possibility that the transit times of Kepler-454b might differ from a linear ephemeris due to light travel time effects or perturbations from the non-transiting companions. Starting with the best-fit solution from the MCMC fit to the short cadence data, we fit each transit event separately allowing the transit center to vary but holding $a/R_\star$, $R_p/R_\star$, and $b$ constant. Although the individual times of transit shifted by $-10$ to $+8$~minutes, these were consistent with the timing precision for individual transits and we did not find evidence for correlated shifts in the transit times.  We therefore adopted a linear ephemeris for Kepler-454b (see Table \ref{table:tparam}).

There is a small possibility that the transit of Kepler-454 is a false positive, and the transit signal is due to an eclipsing binary (either physically bound, or a change alignment on the sky). \citet{mor11} estimate the false positive probability of the Kepler-454 system at about 1\% based on Galactic structure models. \citet{fre13} simulated blends of eclipsing binaries and estimated the fraction that would occur and pass Kepler candidate vetting procedures; those authors estimated a false positive rate of 6.7\% for planets 2-4 $R_{\oplus}$ (appropriate for Kepler-454b). By observing the change in radial velocity of the star at the orbital period and phase expected from the light curve (see Section 5), we rule out the possibility of a false positive scenario.

\section{Radial Velocity Observations \& Reduction}

\subsection{HARPS-N Observations}
We measured the radial velocity (RV) variation of Kepler-454 using the HARPS-N spectrograph on the 3.57m Telescopio Nazionale Galileo (TNG) at the Observatorio del Roque de los Muchachos \citep{cos12}. 
HARPS-N is a highly precise, high-resolution ($R$ $\simeq$ 115,000), vacuum-stabilized spectrograph, very similar in design to the original HARPS planet hunting instrument at the ESO 3.6m \citep{may03}. Notable improvements include the use of octagonal fibers to improve the scrambling of incoming light and a monolithic 4096 x 4096 CCD.

We obtained 55 observations of Kepler-454 during the 2014 observing season and 10 observations during the 2015 season. 
 Most observations were made with a 30 minute exposure time, achieving a mean S/N per pixel of 48 at 5500 \AA\ and a mean internal precision of 2.2 m s$^{-1}$, estimated by combining photon noise, wavelength calibration noise and instrumental drift. Two of these observations had S/N $<$ 30, corresponding to $>$5 m s$^{-1}$ radial velocity precision, and were not included in the analysis. We observed Kepler-454 without simultaneous wavelength reference, to prevent contamination of the stellar spectrum with light from the ThAr calibration lamp.

The spectra were reduced with the standard HARPS-N pipeline and we measured the radial velocities by using a weighted cross-correlation between the observed spectra and a numerical mask based on the spectrum of a G2V star \citep{bar96,pep02}. The resulting radial velocity data are listed in Table \ref{table:rvdat}, with their 1$\sigma$ internal uncertainties, epoch in BJD$_{\rm UTC}$, and bisector span.

\subsection{HIRES Observations}

We obtained 36 observations of Kepler-454 made with the HIRES spectrograph on the Keck Telescope \citep{vog94} through collaboration with the California Planet Search team. These observations were obtained between Aug. 2010 and Dec. 2014, with a mean S/N of 165 and a typical internal precision of 1.4 m s$^{-1}$. The radial velocity measurements were calibrated using an iodine absorption cell \citep{but96}, and their reduction is described in \citet{mar14}. They are listed in Table \ref{table:rvdat2}, with their epoch in BJD$_{\rm UTC}$ and their 1$\sigma$ internal errors. These observations were made as part of NASA's $Kepler$ Key Project follow-up program, and as such all raw and reduced spectra are made available to the public in the Keck Observatory Archive, and the radial velocity measurements will be made available on CFOP.

\section{Analysis of Radial Velocity Measurements}

The combined HARPS-N and HIRES measurements of Kepler-454 show two Keplerian orbits and a long-term trend consistent with an additional companion. We join these data using a single offset term for the HIRES observations and model the radial velocities as a sum of two Keplerian signals, plus a linear trend:
\begin{equation}
\begin{split}
\mathcal{M}(t_{i}) = \gamma\ +\ {\rm RV}_{\rm off} + \beta t_{i} + \\
\sum_{j=1}^{2} K_{j}[\mbox{cos}(\theta_{j}(t_{i},T_{p,j},P_{j},e_{j}) + \omega_{j}) + e_{j} \mbox{cos}(\omega_{j})]
\end{split}
\end{equation}
where RV$_{\rm off}$ is the offset of the HIRES observations from the HARPS-N observations (for the HIRES data, the velocities of each observation are estimated relative to a chosen epoch and hence are relative; for the HARPS-N data, the velocities are measured relative to a theoretical template tied to laboratory rest wavelengths), $\gamma$ is the systemic velocity of Kepler-454 and $\beta$ is the slope of the linear trend due to a long-period companion. Individual orbits $j$ are characterized by their semi-amplitude $K$, period $P$, time of periastron passage $T_{p}$, eccentricity $e$, and argument of periastron $\omega$. The function $\theta$ is the true anomaly of the planet at epoch $t_{i}$. We included the time of reference transit $T_{0}$ as an additional constraint relating $\omega_{j}$, $T_{p,j}$ and $e_{j}$. We use the convention for circular orbits that $\omega = 90^{\circ}$, such that $T_{p}$ = $T_{0}$ and use the standard relationships for eccentric orbits:
\begin{equation}
\theta_{0} = 90^{\circ} - \omega
\end{equation}
\begin{equation}
\mbox{tan}\Bigg(\frac{\theta_{0}}{2}\Bigg) = \Bigg(\frac{1+e}{1-e}\Bigg)^{1/2}\mbox{tan}\Bigg(\frac{E_{0}}{2}\Bigg)
\end{equation}
\begin{equation}
T_{0} - T_{p} = E_{0} - e\mbox{sin}(E_{0}) \frac{P}{2\pi}
\end{equation}
where $\theta_{0}$ is the true anomaly of transit time and $E_{0}$ is the eccentric anomaly of transit time \citep{dan88}. 

We considered circular and eccentric orbits for both planets. We fit an initial solution using a Levenberg-Marquardt minimization algorithm\footnote{https://github.com/pkgw/pwkit} and then these parameters were used as input to \texttt{emcee}, an Affine Invariant MCMC ensemble sampler package \citep{for13}. We initialized 200 chains with starting positions selected by perturbing each free parameter of the LM solution by an amount drawn from a tight Gaussian distribution with width 10$^{-6}$ times the magnitude of that parameter, consistent with the suggested initialization of \texttt{emcee}. 
We set uniform priors on all parameters except $P_{1}$ and $T_{0,1}$, for which we use Gaussian priors determined by the best-fit values from Section 4. We force $K$ to be non-negative and we transformed variables $e$ and $\omega$ to $\sqrt e$ cos$(\omega)$ and $\sqrt e$ sin$(\omega)$ to improve the convergence of low-eccentricity solutions.

As in \citet{dum14}, we model the stellar signal as a constant jitter term $\sigma_{j}$ and use the following likelihood:
\begin{equation}
\mathcal{L} = \prod_{i=1}^{N}\Bigg( \frac{1}{\sqrt{2\pi(\sigma_{i}^{2}+\sigma_{j}^{2})}}\exp[-\frac{(\mbox{RV}(t_{i}) - \mathcal{M}(t_{i}))^{2}}{2(\sigma_{i}^{2}+\sigma_{j}^{2})}]\Bigg)
\end{equation}

where RV$(t_{i})$ is the observed radial velocity at time $t_{i}$ and $\mathcal{M}$ is the model. The stellar jitter noise $\sigma_{j}$ is assumed to be constant and $\sigma_{i}$ is the internal noise for each epoch. The stellar jitter is forced to be positive and allowed to have different values for the HARPS-N and HIRES datasets.


We check for convergence by computing the Gelman-Rubin reduction factor and determined the chains to have converged once all variables had attained \textit{\^R} $<$ 1.03. We discard the first 50\% of each chain as the `burn in' stage and combine the remaining portion of the chains. We select the median value of each parameter as the best-fit value and the error bars are chosen to span the 68\% of values closes to the best-fit value. These values are shown in Table \ref{table:rvparam}. We note that the reported errors on $\gamma$ include only the statistical errors on this fit and not other effects, such as gravitational redshift. The total uncertainty on the systemic velocity is of order 100 m s$^{-1}$.

\begin{figure}
\centering
\includegraphics[width=0.9\textwidth]{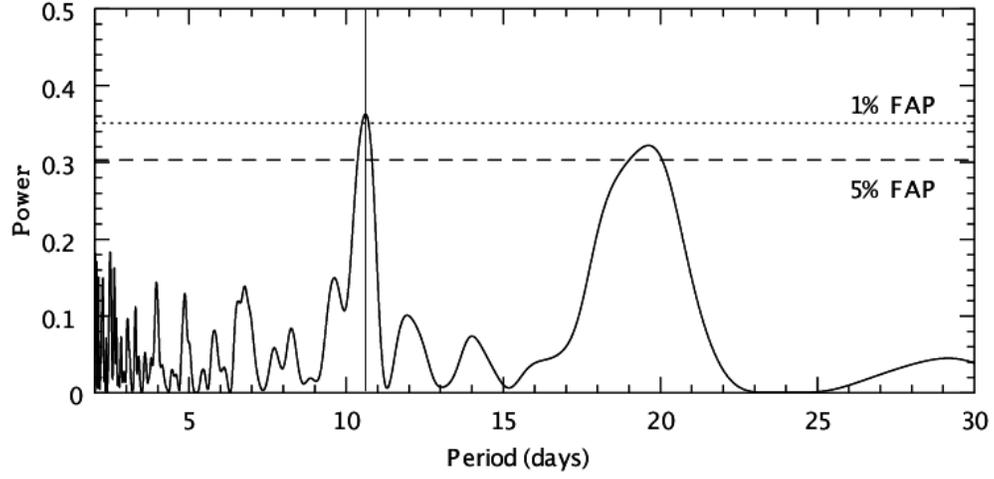}
\caption{Periodogram of the 2014 HARPS-N RV measurements of Kepler-454, after removal of the signal from the outer companions. The vertical line marks the period of Kepler-454b and the 1\% and 5\% FAP levels are shown as dashed lines. \label{rvperio}}
\end{figure}

We consider the null hypothesis that the Kepler-454 system can be described by a linear trend, a marginally eccentric planet in a 524d orbit and Gaussian noise. We compare it to the alternative hypothesis that the Kepler-454 system can be described by those same components, plus a planet in a 10.6d circular orbit. We select the model that best describes the data by considering the Bayesian Information Criterion (BIC).  We use the BIC values to approximate the Bayes factor between pairs of models, with the stellar jitter terms held fixed at 1.6 m s$^{-1}$ and 3.5 m s$^{-1}$ for HARPS-N and HIRES, respectively. These values are representative of those obtained when the jitter terms were allowed to vary. We find a BIC value of 11.9 in favor of including a circular orbit at 10.6 days. Additionally, we show the periodogram of the 2014 HARPS-N radial velocity measurements in Figure \ref{rvperio}, after removing the signal of the outer companions. This series of measurements is most sensitive to short period signals due to its higher cadence and we recover a 10.6d periodicity  with a  false-alarm probability (FAP) of just under 1\%. This is sufficient evidence that the signal of Kepler-454b is present in the radial velocity measurements.

We next consider whether the orbits are sufficiently modeled as sinusoids. We find a BIC value of 6.9 in favor of a circular orbit for Kepler-454b and a BIC value of 0.9 in favor of a marginally eccentric orbit for Kepler-454c. As the BIC values are an approximation, we performed an independent differential evolution MCMC (DE-MCMC) analysis of the combined radial velocity measurements, and found nearly identical values for the median orbital parameters and their error estimates, for both a circular and eccentric fit to the inner planet. In this case, we used 2N chains, where N is the number of free parameters. We imposed Gaussian priors on the period and transit time of Kepler-454b, and Jeffery's priors on the jitter terms. We stopped the chains after they achieved \textit{\^R} $<$ 1.01 and more than 1000 independent draws \citep{for06,bon14}. The Bayes factor values were taken directly from the DE-MCMC posterior distributions by using the Truncated Posterior Mixture method \citep{tuo12}. We estimated a Bayes factor of 39.2 $\pm$ 2.6 in favor of an eccentric solution for Kepler-454c and a Bayes factor of 3.5 $\pm$ 0.3 in favor of an eccentric solution for Kepler-454b. This is strong evidence in favor of an eccentric orbit for Kepler-454c and slight evidence for an eccentric solution for Kepler-454b \citep{kas95}. 

\begin{figure}
\centering
\includegraphics[width=0.9\textwidth]{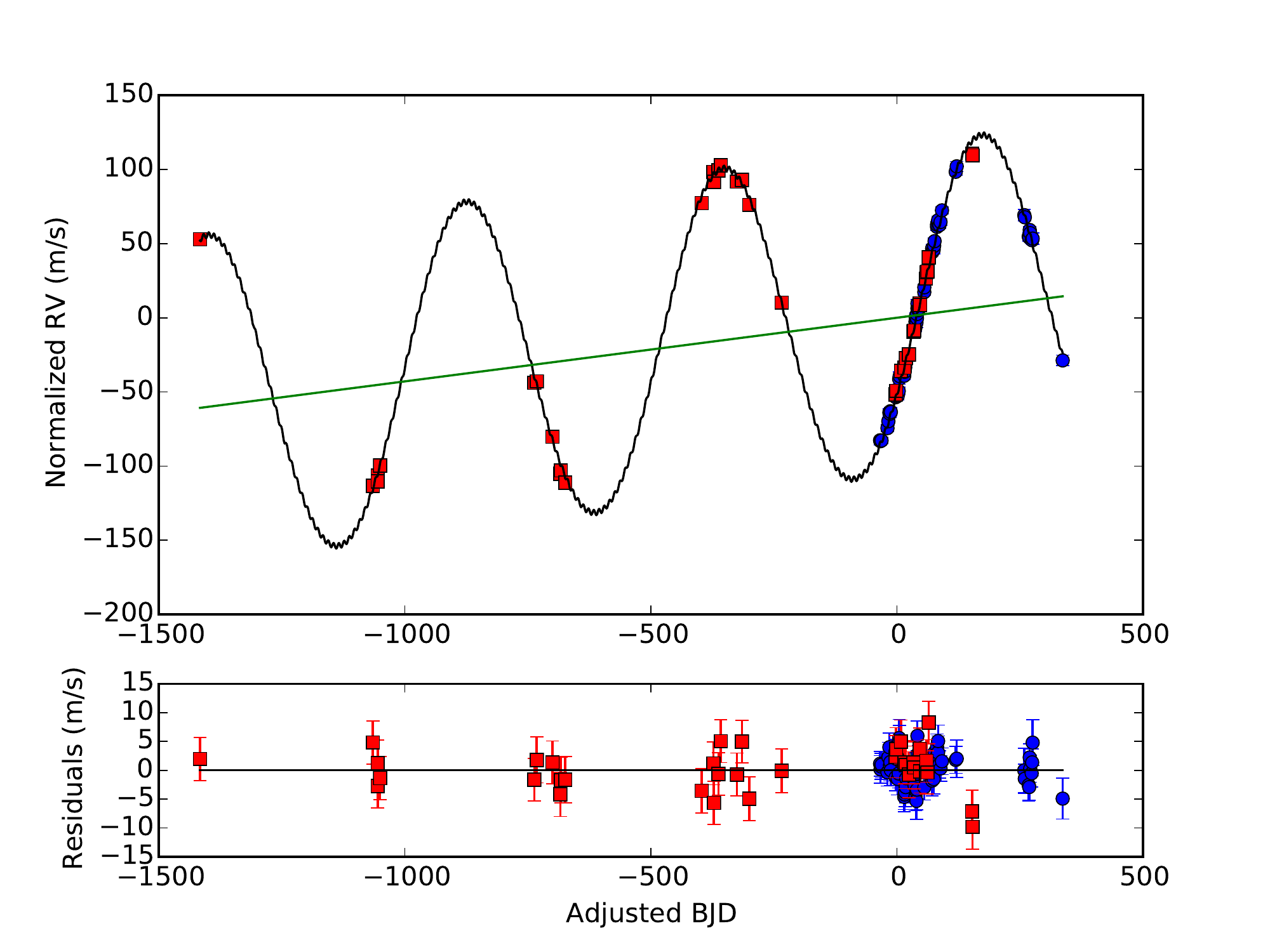}
\caption{\textbf{Top: }RV observations from HARPS-N (blue circles) and Keck-HIRES (red squares), along with the best-fit circular orbit + Keplerian orbit + trend model (black line) and trend only (green line). The error bars are the internal measurement errors and jitter combined in quadrature. \textbf{Bottom: }The residuals to the best-fit model are shown (blue circles \& red squares).\label{rvouter}}
\end{figure}

\begin{figure}
\centering
\includegraphics[width=0.9\textwidth]{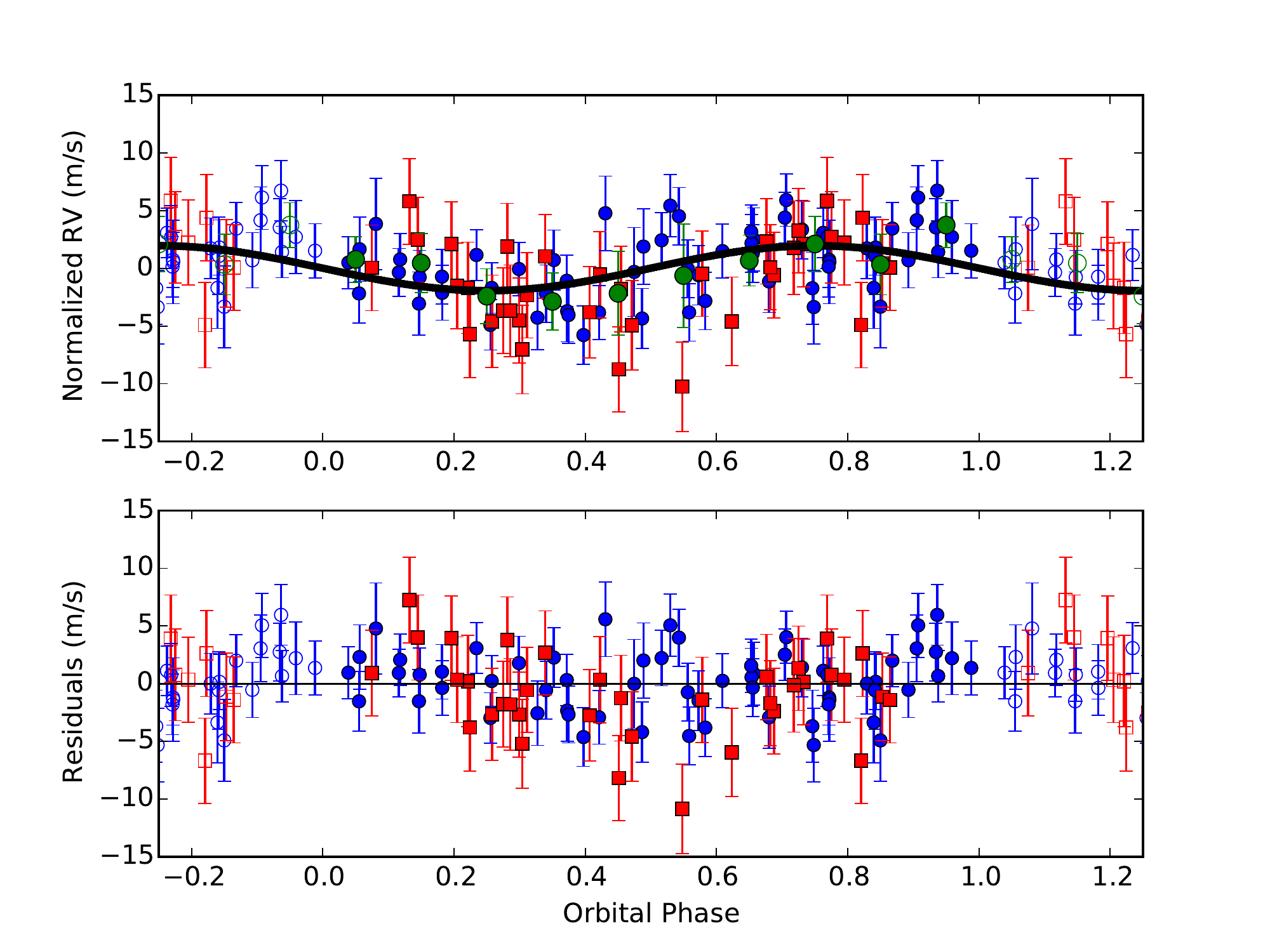}
\caption{\textbf{Top: }Phased RV observations from HARPS-N (blue circles) and Keck-HIRES (red squares) after subtracting the outer orbits. The best-fit circular model for the inner planet is shown as a black line. The error bars are the internal measurement errors and jitter combined in quadrature. The large green circles show the observations binned at intervals of 0.1 orbital phase. Open symbols denote repeated phases. \textbf{Bottom: }The residuals to the best-fit model are shown (blue circles \& red squares). \label{rvinner}}
\end{figure}

\begin{figure}
\centering
\includegraphics[width=0.9\textwidth]{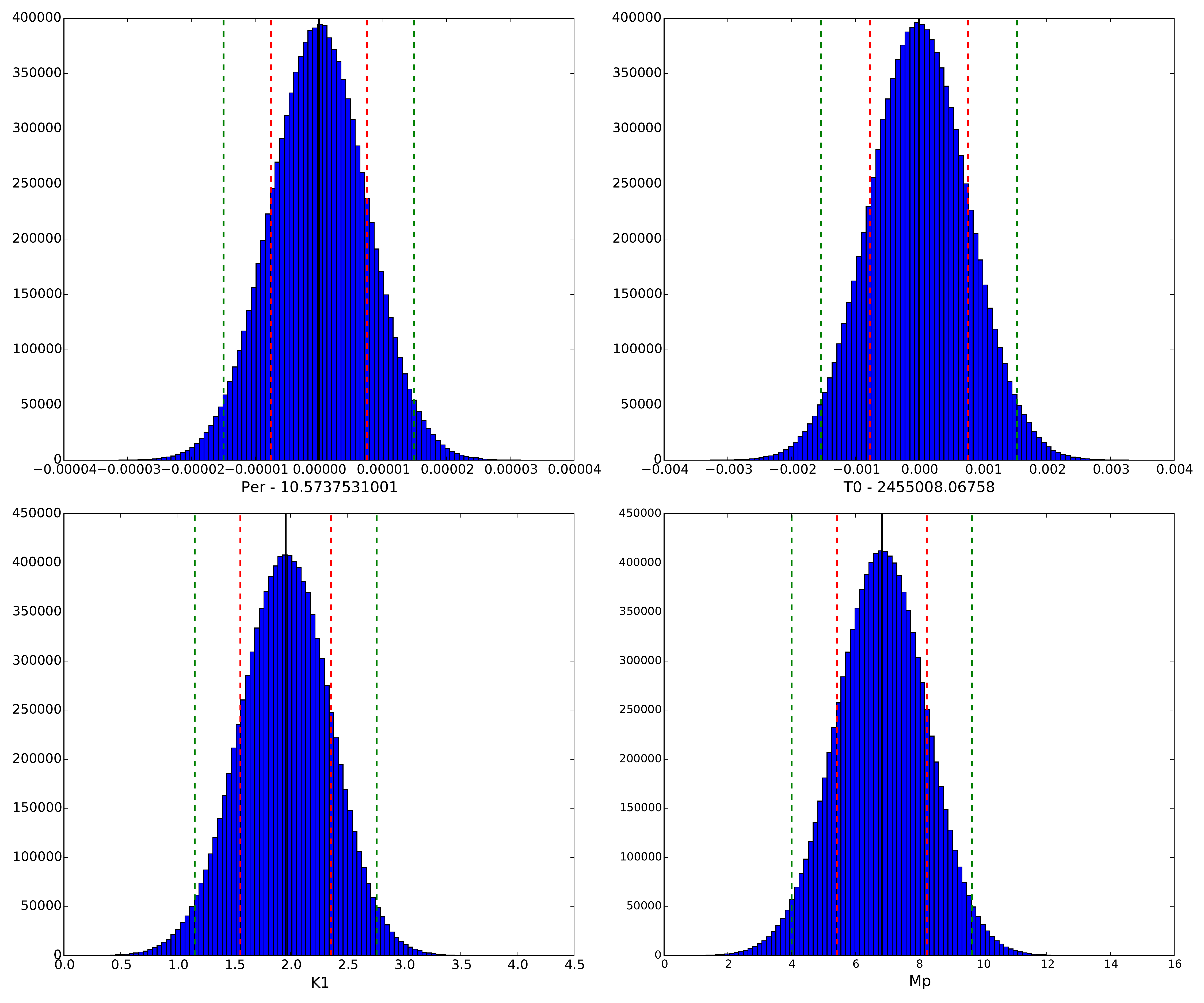}
\caption{The posterior distributions of the parameters in the circular fit to Kepler-454b are shown in blue. The median of the distribution is marked with a black line and the dashed red and green lines contain 68.3\% and 95.4\% of the values, respectively. \label{innercirc_hist}}
\end{figure}

\begin{figure}
\centering
\includegraphics[width=0.9\textwidth]{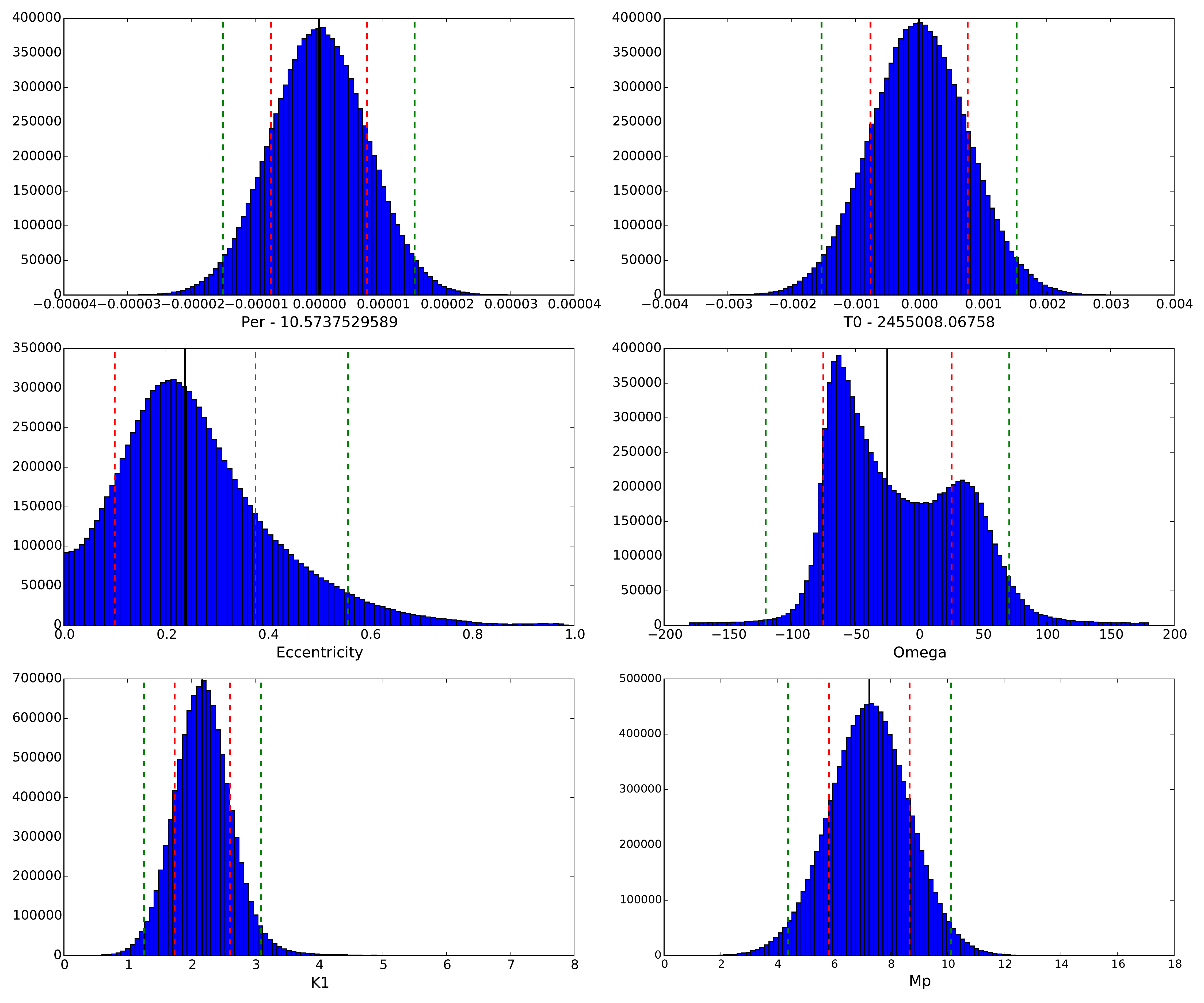}
\caption{The posterior distributions of the parameters in the eccentric fit to Kepler-454b are shown in blue. The median of the distribution is marked with a black line and the dashed red and green lines contain 68.3\% and 95.4\% of the values, respectively. \label{innerecc_hist}}
\end{figure}

\begin{figure}
\centering
\includegraphics[width=0.9\textwidth]{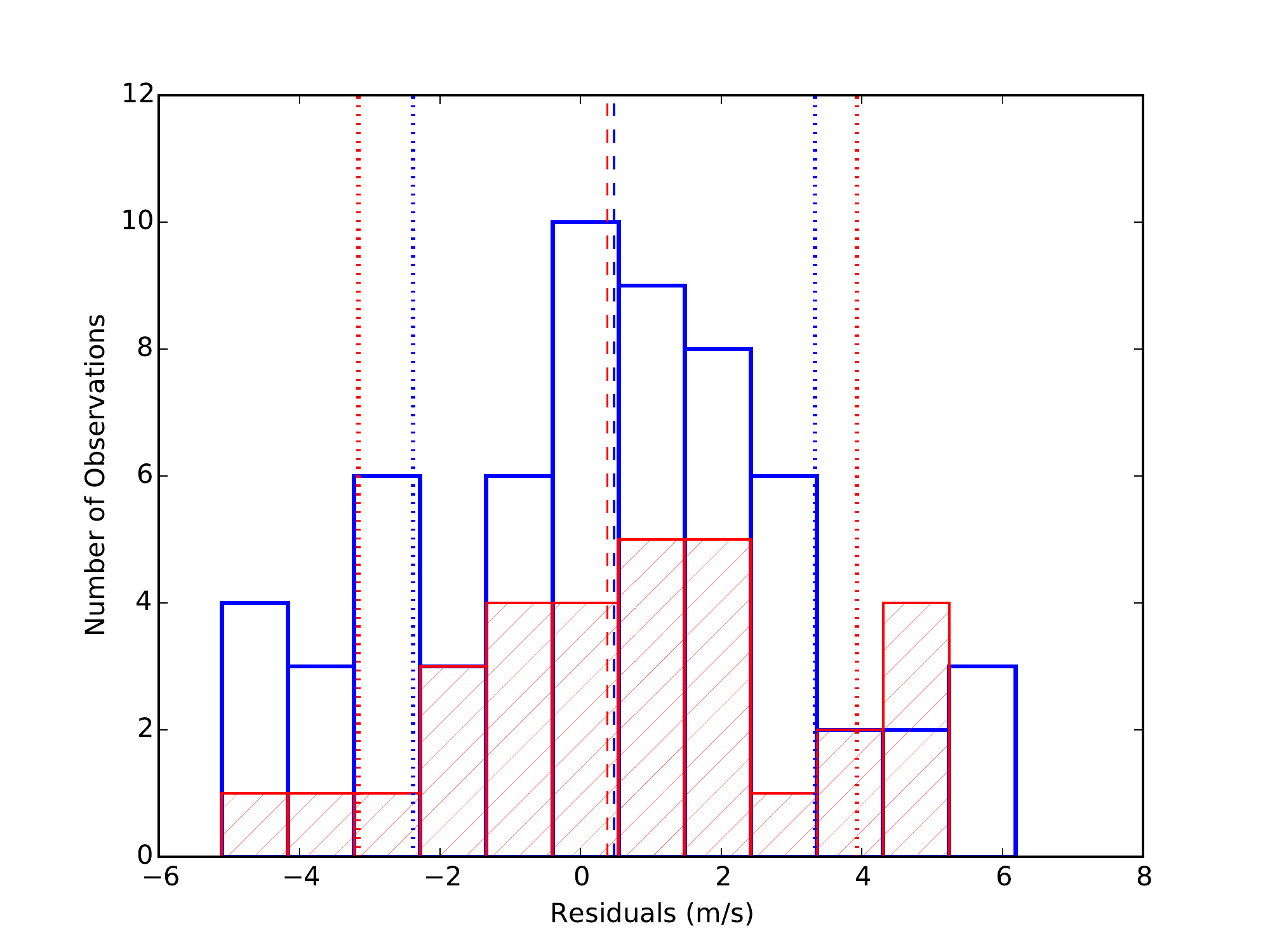}
\caption{Histogram of best-fit residuals to the HARPS-N observations (blue) and HIRES observations (red). The dashed line marks the medians of each distribution, 0.01 m s$^{-1}$ for HARPS-N and 0.36 m s$^{-1}$ for HIRES. The dotted lines mark symmetric error bars including 68\% of the measurements nearest the medians, 2.5 m s$^{-1}$ for HARPS-N and 3.5 m s$^{-1}$ for HIRES.  \label{tel_res}}
\end{figure}

We adopt the simpler model, using an eccentric outer orbit and circular inner orbit for the best-fit model, obtaining a mass estimate for Kepler-454b of 6.84 $\pm$ 1.40 $M_{\oplus}$. The measured RVs and best-fit model are displayed in Figures \ref{rvouter} and \ref{rvinner}. The posterior distributions of the orbital parameters of Kepler-454b are shown in Figure \ref{innercirc_hist}. 
When the eccentricity of the inner planet is allowed to vary, we obtain $e$ = 0.23 $\pm$ 0.13 and a mass estimate of 7.24 $\pm$ 1.40 $M_{\oplus}$, consistent with the results from the best-fit solution. We include the parameters of this solution in Table \ref{table:rvparam} for completeness and show the corresponding posterior distributions in Figure \ref{innerecc_hist}. The distribution of residuals to the best-fit solution are shown in Figure \ref{tel_res}. The HARPS-N residuals have a median value of 0.01 m s$^{-1}$ and 68\% of the values fall within 2.5 m s$^{-1}$ of the median. For the HIRES residuals, the median is 0.36 m s$^{-1}$ and the distribution is a bit broader, with 68\% of the values fall within 3.5 m s$^{-1}$ of the median. 

We calculate the tidal circularization timescale of Kepler-454b using the formula of Goldreich \& Souter (1966) for a 6.84 $M_{\oplus}$, 2.37 $R_{\oplus}$ planet in an 0.095 AU orbit around a 1.028 $M_{\odot}$ star. As Kepler-454b is intermediate between a rocky world and a Neptune-like composition, we expect that its tidal quality factor $Q$ may be intermediate as well. If we assume a $Q$ value of 100, consistent with terrestrial planets in the Solar System, the tidal circularization timescale is $440$ million years, shorter than the 5.2 $\pm$ 1.4 billion year age of the system estimated with asteroseismology. If we assume a $Q$ value of 9000, consistent with Neptune, the circularization timescale is 39 billion years. Obtaining a circularization timescale consistent with the age of Kepler-454 is possible with $Q$ = 1200, a value that is larger than those seen in terrestrial planets, but smaller than that of Neptune \citep{hen09,zha08}.
Though Kepler-454b may have had enough time to reach a circular orbit, the circularization timescale is not sufficiently well determined to discount an eccentric orbit.

As Kepler-454 has both quality asteroseismology and transit data, we are able to calculate a minimum eccentricity for the inner planet through asterodensity profiling in the manner of \citet{kip14}. We compare the stellar density value of 1.199 $\pm$ 0.015 g cm$^{-3}$ obtained through asteroseismology with the stellar density value of 1.04 $\pm$ 0.19 g cm$^{-3}$  implied by a transiting circular orbit and obtain a minimum eccentricity of 0.05 $\pm$ 0.06, consistent with a circular orbit. Given the ambiguity in whether an eccentric model is necessary for Kepler-454b, we favor the simpler solution but provide parameters for both models in Table \ref{table:rvparam}.


Characterization of Kepler-454d is difficult, as the period of its orbit is much longer than the timescale of the combined radial velocity measurements. We observe a linear drift rate of 15.7 $\pm$ 0.6 m s$^{-1}$ yr$^{-1}$ over nearly 5 years. 
Assuming an edge-on circular orbit, this suggests that Kepler-454d has P $>$ 10 yr, $M$sin($i$) $>$ 12.1 $M_{J}$ and a semi-major axis $>4.7$ AU. Assuming that this solar-like star has an absolute magnitude $M_{V}=4.9$, comparable to the Sun, its distance is approximately 200 pc and the angular separation at maximum elongation of Kepler-454d is $>$0\farcs{02}. 

\begin{figure}
\centering
\includegraphics[width=0.9\textwidth]{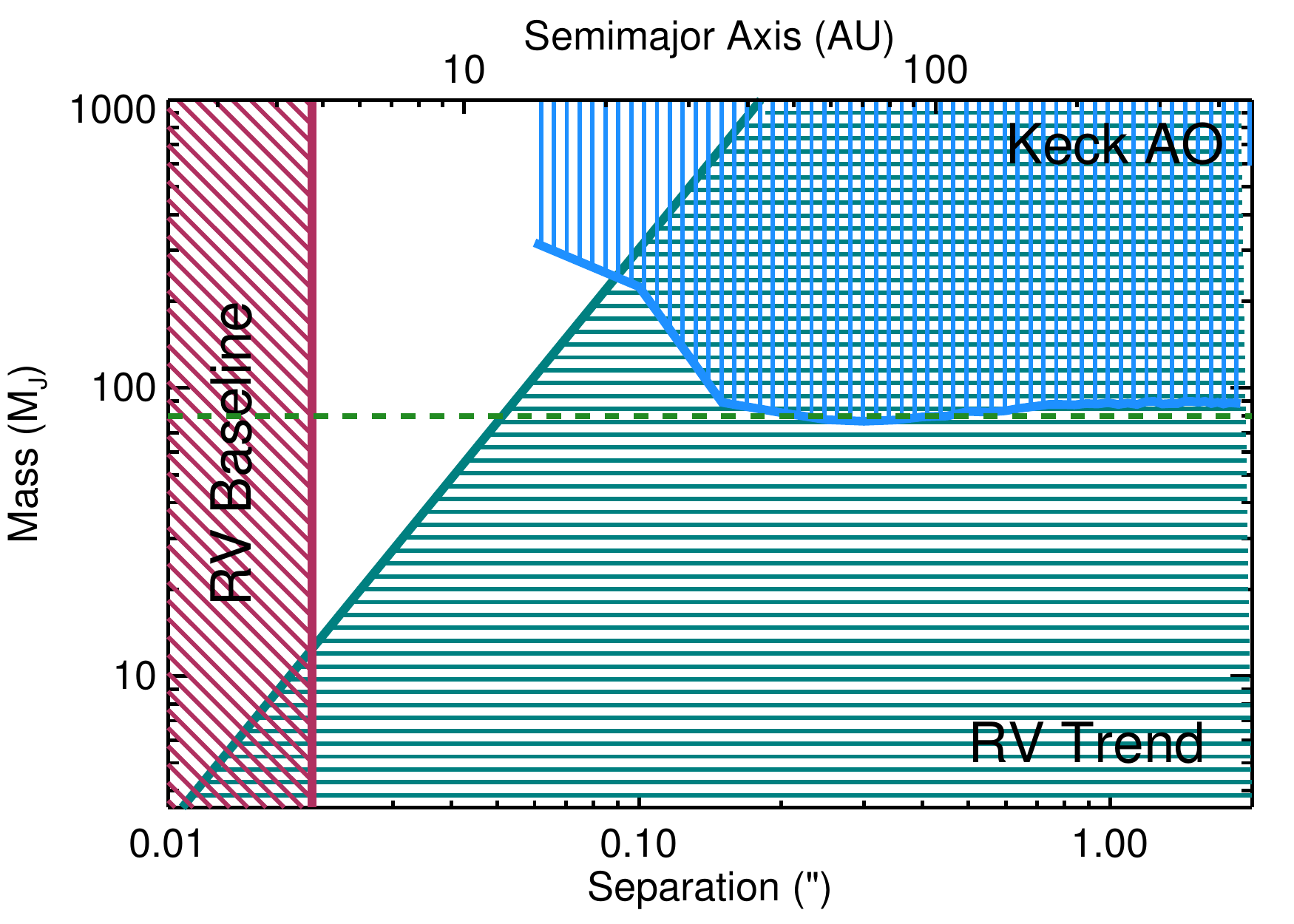}
\caption{Limits on the mass and separation of Kepler-454d. The baseline of the combined RV observations excludes the maroon region and the amplitude of the RV trend excludes the teal region. The blue region is excluded by the Keck AO observations. These limits assume a circular, coplanar orbit for Kepler-454d. The dashed green line marks 80 $M_{J}$, the mass boundary between brown dwarfs and stars. \label{ao_rv_lim}}
\end{figure}

There are multiple sources of adaptive optics (AO) observations for this target, though none are able to place limits on the brightness of the companion at such a small separation. The most stringent of these are Keck observations made in $Br_{\gamma}$, with a brightness limit at 0\farcs{06} of 3.73 magnitudes fainter than the host star (CFOP; D. Ciardi). Assuming that $K_{s}$ magnitude is equivalent to $Br_{\gamma}$ magnitude, Kepler-454d must then be fainter than $K_{s}$ = 13.7, unless it was not detectable at the time of the AO observations due to orbital geometry. We converted the $K_{s}$ limit into a mass upper limit using the \citet{del00} relation and found a value of 300 $M_{J}$ for angular separations beyond 0\farcs{05}. The combined restrictions from the RV and AO data are shown in Figure \ref{ao_rv_lim}. 

Astrometry from the Gaia mission could readily complement our radial velocity measurements. It will provide approximately 16 $\mu$as astrometry for stars with $V$ = 12 mag \citep{els14}. With this precision, Jupiter-sized planet like Kepler-454c would be marginally detectable as a signal of order 30 $\mu$as, providing measurements of inclination and mass.  The acceleration due to the outer companion should be readily detectable, further constraining its separation and mass.

\begin{figure}
\centering
\includegraphics[width=0.9\textwidth]{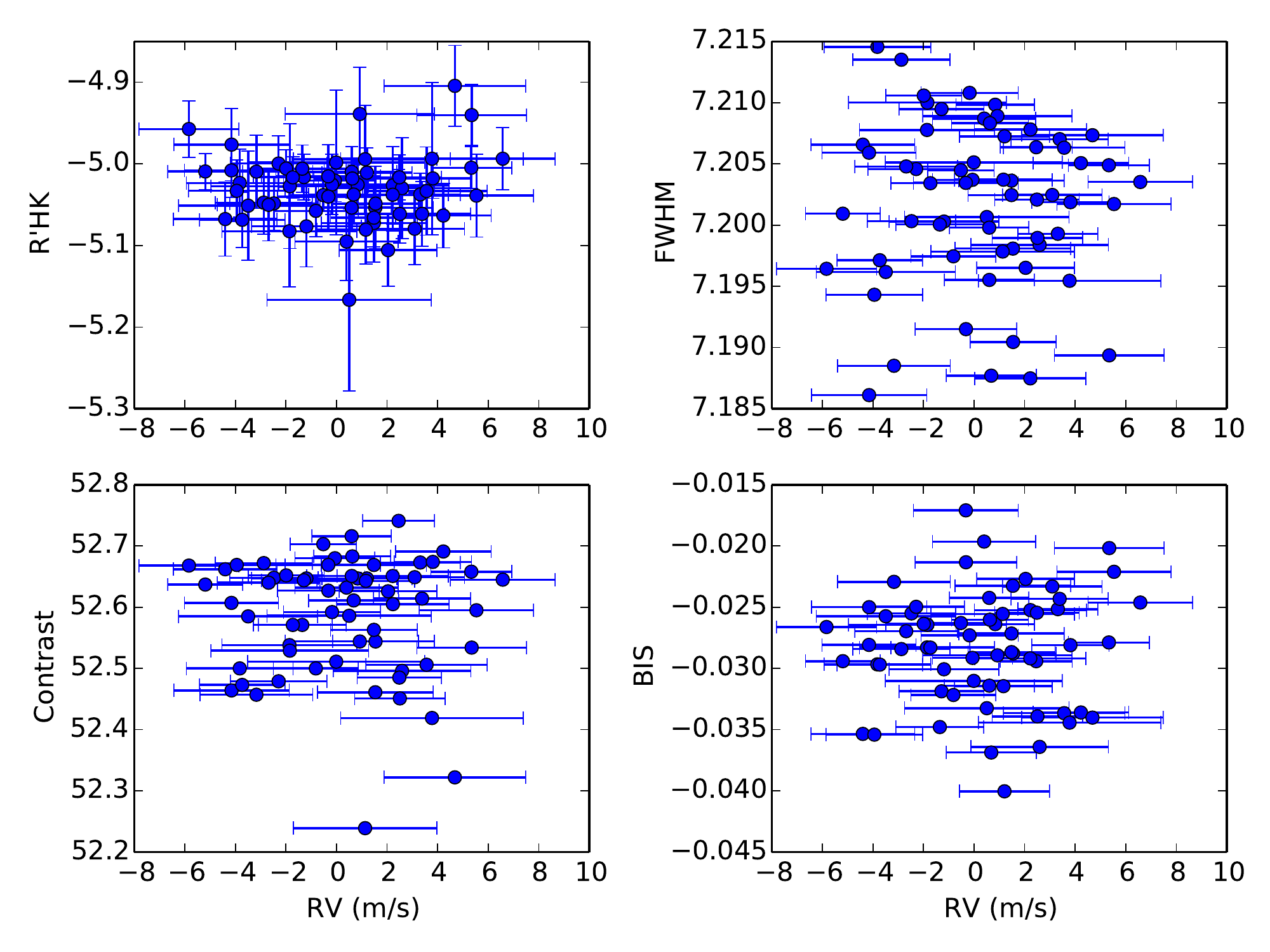}
\caption{Plots of the log $R'_{\rm HK}$ activity index and the FWHM, contrast and bisector velocity span of the cross-correlation function as a function of the de-trended radial velocity measurements. No correlations are found. \label{activity}}
\end{figure}

While Kepler-454 is an inactive star, with a median log $R'_{\rm HK}$ value of -5.0, we consider the possibility of radial velocity variations induced by stellar activity. After removing the signal from the 524 d orbit and linear trend, we compare the radial velocity measurements to the log $R'_{\rm HK}$ values, and several features of the cross-correlation function (CCF), including the bisector velocity span (BIS), FWHM and the contrast of the CCF. We found no correlations with radial velocity, as shown in Figure \ref{activity} and there are no significant periodicities in these indicators. We conclude that it is sufficient to assume a Gaussian noise term in the likelihood function, to model RV variations due to stellar activity.

\section{Discussion}
We present a mass measurement for Kepler-454b of 6.8 $\pm$ 1.4 $M_{\oplus}$ and detect two additional non-transiting companions, a planet with a minimum mass of 4.46 $\pm$ 0.12 $M_{J}$ in a slightly eccentric 524 d orbit and a linear trend consistent with a brown dwarf or low-mass star. 

\begin{figure}
\centering
\includegraphics[width=1.0\textwidth]{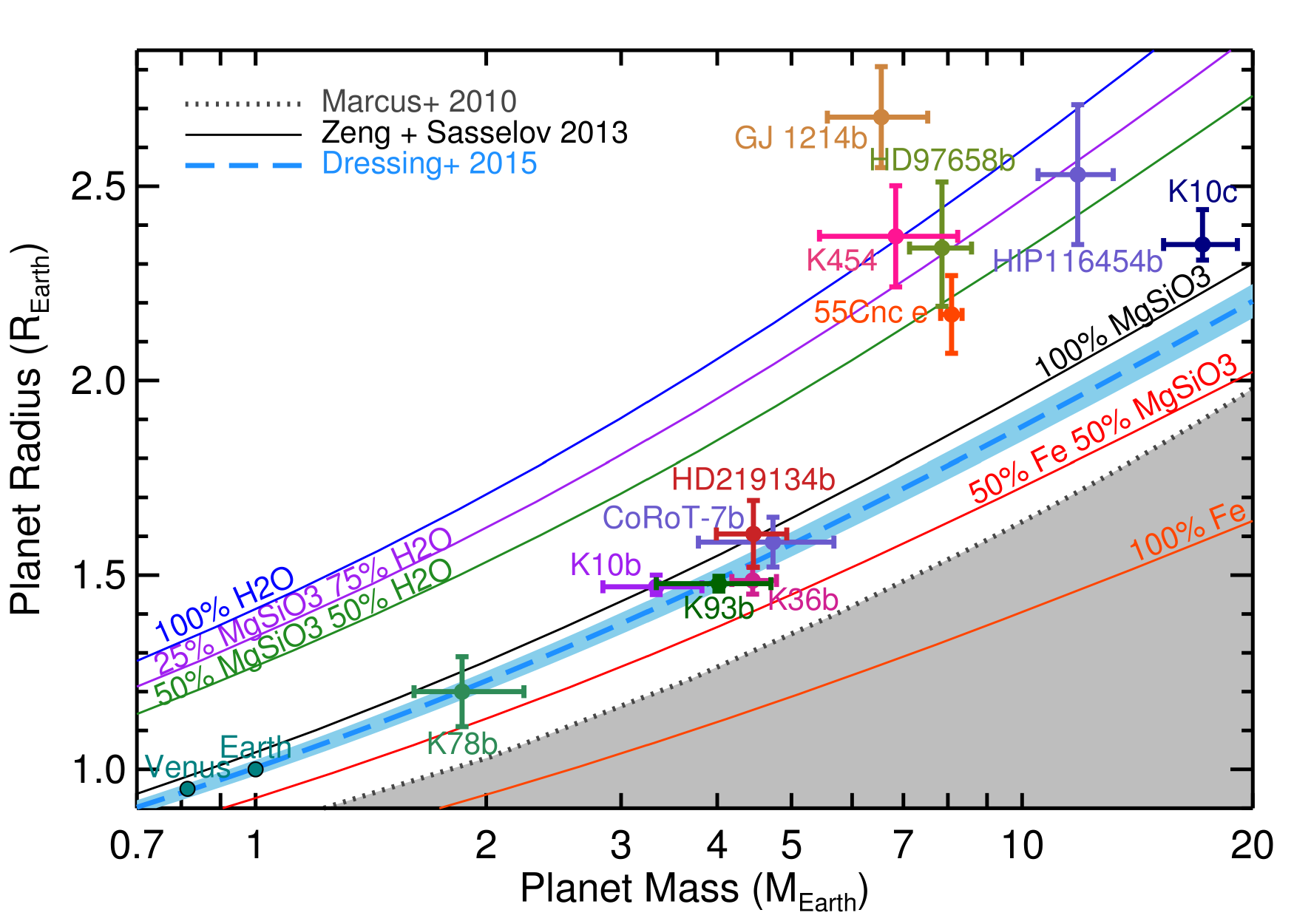}
\caption{Mass-radius diagram for planets with radii $<$2.7 $R_{\oplus}$ and masses measured to better than 20\% precision. The shaded gray region in the lower right indicates planets with iron content exceeding the maximum value predicted from models of collisional stripping \citep{mar10}. The solid lines are theoretical mass-radius curves \citep{zen13} for planets with compositions of 100\% H$_{2}$O (blue), 25\% MgSiO$_{3}$ - 75\% H$_{2}$O (purple), 50\% MgSiO$_{3}$ - 50\% H$_{2}$O (green), 100\% MgSiO$_{3}$ (black), 50\% Fe - 50\% MgSiO$_{3}$ (red), and 100\% Fe (orange). Our best-fit relation based on the \citet{zen13} models is the dashed light blue line representing an Earth-like composition (modeled as 17\% Fe and 83\% MgSiO$_{3}$ using a fully differentiated, two-component model). The shaded region surrounding the line indicates the 2\% dispersion in the radius expected from the variation in Mg/Si and Fe/Si ratios \citep{gra09}.\label{mrplot}}
\end{figure}

Combining our mass estimate for Kepler-454b with a radius of 2.37 $\pm$ 0.13 $R_{\oplus}$, gives a density estimate of 2.76 $\pm$ 0.73 g cm$^{-3}$. Figure \ref{mrplot} shows a mass-radius plot of Kepler-454b along with the several other planets smaller than 2.7 $R_{\oplus}$ and with masses measured to better than 20\% precision. There are twelve such planets including Kepler-454b. \citet{dre15b} note that the six planets with radii $<$1.6 $R_{\oplus}$, as well as Earth and Venus, have very similar uncompressed densities. The recently detected planet HD 219314b also has a comparable density \citep{mot15}. These planets are consistent with an Earth-like composition, notably the Earth's ratio of iron to magnesium silicates. 
In contrast, the six planets with radii $2.0\leq$ $R$ $(R_{\oplus})$ $\leq 2.7$ are not consistent with this rocky composition model. Their lower densities require a significant fraction of volatiles, likely in the form of an envelope of water and other volatiles and/or H/He. If rocky planets with radii $>$1.6 $R_{\oplus}$ do exist, they are likely to be more massive and thus easier to detect. If such high density planets continue to be absent as the sample of small planets grows, it would suggest that most planets with masses greater than about 6 $M_{\oplus}$ may contain a significant fraction of volatiles and/or H/He. 

Kepler-454 has similar parameters to the Sun and does not appear to be unique compared to the other host stars in this sample.  The transit parameters initially implied by CFOP placed Kepler-454b in a potentially unique space in the mass-radius diagram, with a radius estimate between the population of planets with Earth-like densities and the population of larger, less dense planets. With its radius measurement now revised upward due to analysis of the short cadence observation, Kepler-454b has a both mass and radius comparable to several of the less dense planets. At 2.76 $\pm$ 0.73 g cm$^{-3}$, it falls well above the Earth-like composition curve and it likely requires a significant fraction of volatiles. 

Specifically, a planet with an Earth-like composition and the same mass as Kepler-454b would have a radius of  $R_{p, Earth-like} = 1.73$ $R_{\oplus}$, significantly smaller than the observed radius of Kepler-454b ($R_{p, obs} = $2.37 $\pm$ 0.13 $R_{\oplus}$). The observed ``radius excess'' $\Delta R_p = R_{p, Earth-like} - R_{p,obs}$ for Kepler-454b is therefore $\Delta R_p = 0.64$ $ R_{\oplus}$.

We can estimate the amount of lower density material required to explain the observed radius of Kepler-454b by assuming that Kepler-454b is an Earth-like mixture of rock and iron covered by a low density envelope. Employing the models of \citet{lop14} and assuming a system age of 5 Gyr, the observed mass and radius of Kepler-454b could be explained if the planet is shrouded by a solar metallicity H/He and a total mass equal to roughly 1\% of the total planetary mass.

\begin{figure*}
\centering
\includegraphics[width=1.0\textwidth]{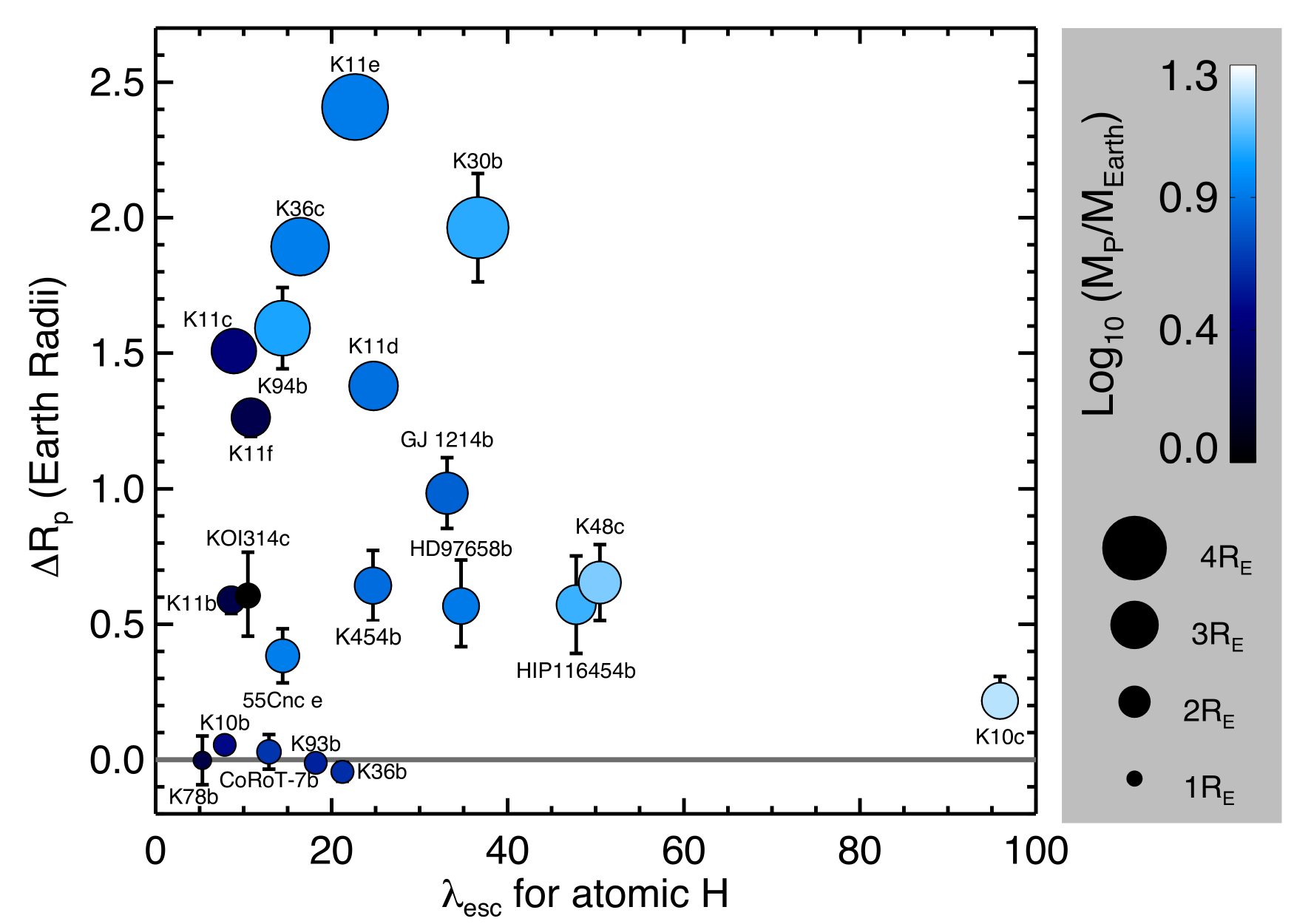}
\caption{Difference between observed planet radius and the radius that would have been expected if the planet had an Earth-like composition versus the Jeans escape parameter. Planets consistent with Earth-like compositions are concentrated near the gray line at $\Delta R_p = 0$. The sizes of the circles are scaled based on the radii of the planets and the colors indicate the planet masses.\label{fig:excessrp}}
\end{figure*}

\begin{figure*}
\centering
\includegraphics[width=1.0\textwidth]{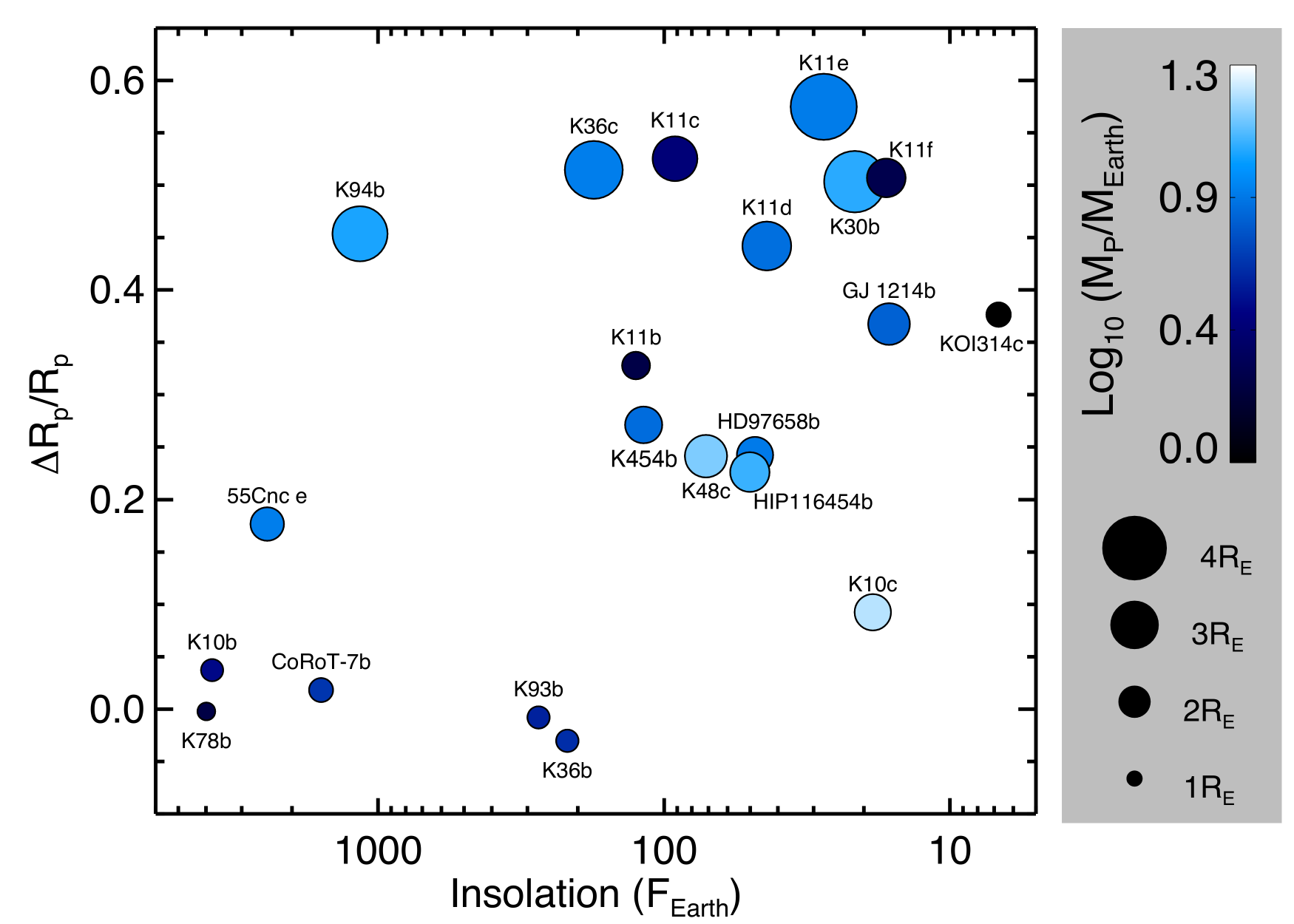}
\caption{The relative radius excess ($\Delta R_p / R_p$) versus the insolation flux received by each planet. As in Figure \ref{fig:excessrp}, the sizes and colors of the circles indicate the radii and masses of the planets, respectively. \label{fig:excessrp2}}
\end{figure*}

In Figure \ref{fig:excessrp}, we compare the observed radius excess $\Delta R_p = R_{p, obs} - R_{p, Earth-like}$ for Kepler-454b to the $\Delta R_p$ estimated for other small transiting planets with mass estimates. The left panel of Figure~\ref{fig:excessrp} displays the $\Delta R_p$ as a function of the Jeans escape parameter $\lambda_{\rm esc}$:
\begin{equation}
\lambda_{\rm esc} \equiv \frac{G M_p m}{k T r_c} 
\end{equation}
where $G$ is the gravitational constant, $M_p$ is the mass of the planet, $m$ is the mean molecular or atomic weight of the atmosphere (here we set $m$ to the value of atomic hydrogen), $k$ is the Boltzmann constant, $r_c$ is the height above the center of the planet and $T$ is the temperature of the exobase, the atmospheric boundary above which particles are gravitationally bound to the planet but move on collision-free trajectories \citep{meadows+seager2010}. We assumed that the temperatures $T$ of the exobase are equal to the expected equilibrium temperatures of the planets for an albedo of 0, but the true exobase temperatures are likely higher. For highly irradiated planets, the dominant atmospheric loss channel is likely hydrodynamic escape rather than Jeans escape. Accordingly, Figure~\ref{fig:excessrp2} presents the ratio $\Delta R_p/R_p$ versus the insolation received by each planet. Compared to the other planets, Kepler-454b is most similar to HD~97658b, HIP~116454b, Kepler~48c, and Kepler-11b. All of these planets receive roughly $100\times$ the insolation received by the Earth and have relative radius excesses $\Delta R_p / R_p$ of approximately $20\%$, consistent with prior work by \citet{lop12} and \citet{owe12}.

For H/He dominated atmospheric envelopes, $\Delta R_p/R_p$ is roughly equivalent to the relative mass fraction of envelope \citep{lop14}.
In general, we find that the planets with lower relative envelope fractions are more highly irradiated than the planets with large relative envelope fractions. However, accurately constraining the masses and radii of small planets becomes increasingly difficult as the orbital period increases. The relative lack of small dense planets receiving low insolation fluxes may therefore be due to an observational bias rather than a real scarcity of cool dense small planets. As the precision of radial velocity spectrographs improves, we may discover additional planets that are smaller, denser, and cooler than Kepler-10c. The NASA TESS Mission, scheduled for launch in 2017, will help by providing hundreds of Earths, super-Earths, and mini-Neptunes transiting stars that are generally much brighter than those from Kepler \citep{sul15}, greatly facilitating RV follow-up and permitting masses to be measured for much longer orbital periods.

In contrast, obtaining a precise mass measurement of a highly irradiated planet with a large relative envelope fraction is observationally easier than measuring the mass of a less strongly irradiated large planet. Accordingly, the relative dearth of highly irradiated small planets with large envelope fractions likely indicates that such planets are rare. As the number of small planets with well-measured masses and radii continues to grow, we will be able to further investigate the properties and the formation, and subsequently loss or retention of gaseous envelopes of small planets.

The authors would like to thank the TNG observers who contributed to the measurements reported here, including Walter Boschin, Massimo Cecconi, Vania Lorenzi and Marco Pedani. We also thank Lauren Weiss for gathering some of the HIRES data presented here. The authors wish to thank the
entire \emph{Kepler} team, without whom these results would not be
possible. Funding for this Discovery mission is provided by
NASA's Science Mission Directorate.
The HARPS-N project was funded by the Prodex Program of the Swiss Space Office (SSO), the Harvard University Origin of Life Initiative (HUOLI), the Scottish Universities Physics Alliance (SUPA), the University of Geneva, the Smithsonian Astrophysical Observatory (SAO), and the Italian National Astrophysical Institute (INAF), University of St. Andrews, Queen's University Belfast and University of Edinburgh. The research leading to these results has received funding from
the European Union Seventh Framework Programme (FP7/2007-2013) under Grant Agreement No. 313014 (ETAEARTH). This publication was made possible by a grant from the John Templeton Foundation. The opinions expressed in this publication are those of the authors and do not necessarily reflect the views of the John Templeton Foundation. This material is based upon work supported by the National Aeronautics and Space Administration under Grant No. NNX15AC90G issued through the Exoplanets Research Program. C. D. is supported by a National Science Foundation Graduate Research Fellowship. X. D. would like to thank the Swiss National Science Foundation (SNSF) for its support through an Early Postdoc Mobility fellowship. A. V. is supported by the National Science Foundation Graduate Research Fellowship, Grant No. DGE 1144152. P. F. acknowledges support by Funda\c c\~ao para a Ci\^encia e a Tecnologia (FCT) through Investigador FCT contracts of reference IF/01037/2013 and POPH/FSE (EC) by FEDER funding through the program ``Programa Operacional de Factores de Competitividade - COMPETE''. PF further acknowledges support from Funda\c{c}\~ao para a Ci\^encia e a Tecnologia (FCT) in the form of an exploratory project of reference IF/01037/2013CP1191/CT0001. W.J.C., T.L.C. and G.R.D. acknowledge the support of the UK
Science and Technology Facilities Council (STFC).  S.B. acknowledges
partial support from NSF grant AST-1105930 and NASA grant NNX13AE70G.
T.S.M. was supported by NASA grant NNX13AE91G. Computational time on
Stampede at the Texas Advanced Computing Center was provided through
XSEDE allocation TG-AST090107. Funding for the Stellar Astrophysics
Centre is provided by The Danish National Research Foundation (Grant
agreement no.: DNRF106). The research is supported by the ASTERISK
project (ASTERoseismic Investigations with SONG and Kepler) funded by
the European Research Council (Grant agreement no.: 267864); and by
the European Community's Seventh Framework Programme (FP7/2007-2013)
under grant agreement no. 312844 (SPACEINN). Partial support was received from the $Kepler$ mission under NASA Cooperation Agreement NNX13AB58A to the Smithsonian Astrophysical Observatory, DWL PI.

Some of the data presented herein were obtained at the W. M. Keck Observatory, which is operated as a  scientific partnership among the California Institute of Technology,  the University of California, and the National Aeronautics and Space  Administration. The Keck Observatory was made possible by the  generous financial support of the W. M. Keck Foundation.   The spectra and their  products are made available at the NExSci Exoplanet Archive and its  CFOP website: {\url{http://exoplanetarchive.ipac.caltech.edu}}.  We  thank the many observers who contributed to the HIRES measurements  reported here, including Benjamin J. Fulton, Evan Sinukoff and Lea Hirsch.  We gratefully acknowledge the efforts and dedication  of the Keck Observatory staff, especially Greg Doppmann, Scott Dahm, Hien Tran, and  Grant Hill for support of HIRES and Greg Wirth and Bob Goodrich for support of remote
  observing.   This research has made use of the NASA Exoplanet Archive, which is operated by the California Institute of Technology, under contract with the National Aeronautics and Space Administration under the Exoplanet Exploration Program.  Finally, the authors wish to extend special thanks to those of Hawai`ian ancestry on whose sacred mountain of Mauna Kea we are privileged to be guests.  Without their generous hospitality, the Keck observations presented herein would not
have been possible.

\clearpage
\bibliography{koi273}

\begin{thebibliography}{}
\expandafter\ifx\csname natexlab\endcsname\relax\def\natexlab#1{#1}\fi

\bibitem[{{Baranne} {et~al.}(1996){Baranne}, {Queloz}, {Mayor}, {Adrianzyk},
  {Knispel}, {Kohler}, {Lacroix}, {Meunier}, {Rimbaud}, \& {Vin}}]{bar96}
{Baranne}, A., {Queloz}, D., {Mayor}, M., {et~al.} 1996, \aaps, 119, 373

\bibitem[{{Batalha} {et~al.}(2013){Batalha}, {Rowe}, {Bryson}, {Barclay},
  {Burke}, {Caldwell}, {Christiansen}, {Mullally}, {Thompson}, {Brown},
  {Dupree}, {Fabrycky}, {Ford}, {Fortney}, {Gilliland}, {Isaacson}, {Latham},
  {Marcy}, {Quinn}, {Ragozzine}, {Shporer}, {Borucki}, {Ciardi}, {Gautier},
  {Haas}, {Jenkins}, {Koch}, {Lissauer}, {Rapin}, {Basri}, {Boss}, {Buchhave},
  {Carter}, {Charbonneau}, {Christensen-Dalsgaard}, {Clarke}, {Cochran},
  {Demory}, {Desert}, {Devore}, {Doyle}, {Esquerdo}, {Everett}, {Fressin},
  {Geary}, {Girouard}, {Gould}, {Hall}, {Holman}, {Howard}, {Howell},
  {Ibrahim}, {Kinemuchi}, {Kjeldsen}, {Klaus}, {Li}, {Lucas}, {Meibom},
  {Morris}, {Pr{\v s}a}, {Quintana}, {Sanderfer}, {Sasselov}, {Seader},
  {Smith}, {Steffen}, {Still}, {Stumpe}, {Tarter}, {Tenenbaum}, {Torres},
  {Twicken}, {Uddin}, {Van Cleve}, {Walkowicz}, \& {Welsh}}]{bat13}
{Batalha}, N.~M., {Rowe}, J.~F., {Bryson}, S.~T., {et~al.} 2013, \apjs, 204, 24

\bibitem[{{Bedding} \& {Kjeldsen}(2010)}]{bed10}
{Bedding}, T.~R., \& {Kjeldsen}, H. 2010, Communications in Asteroseismology,
  161, 3

\bibitem[{{Bonomo} {et~al.}(2014){Bonomo}, {Sozzetti}, {Lovis}, {Malavolta},
  {Rice}, {Buchhave}, {Sasselov}, {Cameron}, {Latham}, {Molinari}, {Pepe},
  {Udry}, {Affer}, {Charbonneau}, {Cosentino}, {Dressing}, {Dumusque},
  {Figueira}, {Fiorenzano}, {Gettel}, {Harutyunyan}, {Haywood}, {Horne},
  {Lopez-Morales}, {Mayor}, {Micela}, {Motalebi}, {Nascimbeni}, {Phillips},
  {Piotto}, {Pollacco}, {Queloz}, {S{\'e}gransan}, {Szentgyorgyi}, \&
  {Watson}}]{bon14}
{Bonomo}, A.~S., {Sozzetti}, A., {Lovis}, C., {et~al.} 2014, \aap, 572, A2

\bibitem[{{Borucki} {et~al.}(2011){Borucki}, {Koch}, {Basri}, {Batalha},
  {Brown}, {Bryson}, {Caldwell}, {Christensen-Dalsgaard}, {Cochran}, {DeVore},
  {Dunham}, {Gautier}, {Geary}, {Gilliland}, {Gould}, {Howell}, {Jenkins},
  {Latham}, {Lissauer}, {Marcy}, {Rowe}, {Sasselov}, {Boss}, {Charbonneau},
  {Ciardi}, {Doyle}, {Dupree}, {Ford}, {Fortney}, {Holman}, {Seager},
  {Steffen}, {Tarter}, {Welsh}, {Allen}, {Buchhave}, {Christiansen}, {Clarke},
  {Das}, {D{\'e}sert}, {Endl}, {Fabrycky}, {Fressin}, {Haas}, {Horch},
  {Howard}, {Isaacson}, {Kjeldsen}, {Kolodziejczak}, {Kulesa}, {Li}, {Lucas},
  {Machalek}, {McCarthy}, {MacQueen}, {Meibom}, {Miquel}, {Prsa}, {Quinn},
  {Quintana}, {Ragozzine}, {Sherry}, {Shporer}, {Tenenbaum}, {Torres},
  {Twicken}, {Van Cleve}, {Walkowicz}, {Witteborn}, \& {Still}}]{bor11}
{Borucki}, W.~J., {Koch}, D.~G., {Basri}, G., {et~al.} 2011, \apj, 736, 19

\bibitem[{{Buchhave} {et~al.}(2012){Buchhave}, {Latham}, {Johansen},
  {Bizzarro}, {Torres}, {Rowe}, {Batalha}, {Borucki}, {Brugamyer}, {Caldwell},
  {Bryson}, {Ciardi}, {Cochran}, {Endl}, {Esquerdo}, {Ford}, {Geary},
  {Gilliland}, {Hansen}, {Isaacson}, {Laird}, {Lucas}, {Marcy}, {Morse},
  {Robertson}, {Shporer}, {Stefanik}, {Still}, \& {Quinn}}]{buc12}
{Buchhave}, L.~A., {Latham}, D.~W., {Johansen}, A., {et~al.} 2012, \nat, 486,
  375

\bibitem[{{Buchhave} {et~al.}(2014){Buchhave}, {Bizzarro}, {Latham},
  {Sasselov}, {Cochran}, {Endl}, {Isaacson}, {Juncher}, \& {Marcy}}]{buc14}
{Buchhave}, L.~A., {Bizzarro}, M., {Latham}, D.~W., {et~al.} 2014, \nat, 509,
  593

\bibitem[{{Burke} {et~al.}(2014){Burke}, {Bryson}, {Mullally}, {Rowe},
  {Christiansen}, {Thompson}, {Coughlin}, {Haas}, {Batalha}, {Caldwell},
  {Jenkins}, {Still}, {Barclay}, {Borucki}, {Chaplin}, {Ciardi}, {Clarke},
  {Cochran}, {Demory}, {Esquerdo}, {Gautier}, {Gilliland}, {Girouard}, {Havel},
  {Henze}, {Howell}, {Huber}, {Latham}, {Li}, {Morehead}, {Morton}, {Pepper},
  {Quintana}, {Ragozzine}, {Seader}, {Shah}, {Shporer}, {Tenenbaum}, {Twicken},
  \& {Wolfgang}}]{bur14}
{Burke}, C.~J., {Bryson}, S.~T., {Mullally}, F., {et~al.} 2014, \apjs, 210, 19

\bibitem[{{Butler} {et~al.}(1996){Butler}, {Marcy}, {Williams}, {McCarthy},
  {Dosanjh}, \& {Vogt}}]{but96}
{Butler}, R.~P., {Marcy}, G.~W., {Williams}, E., {et~al.} 1996, \pasp, 108, 500

\bibitem[{{Christensen-Dalsgaard}(2008)}]{chr08}
{Christensen-Dalsgaard}, J. 2008, \apss, 316, 113

\bibitem[{{Cosentino} {et~al.}(2012){Cosentino}, {Lovis}, {Pepe}, {Collier
  Cameron}, {Latham}, {Molinari}, {Udry}, {Bezawada}, {Black}, {Born},
  {Buchschacher}, {Charbonneau}, {Figueira}, {Fleury}, {Galli}, {Gallie},
  {Gao}, {Ghedina}, {Gonzalez}, {Gonzalez}, {Guerra}, {Henry}, {Horne},
  {Hughes}, {Kelly}, {Lodi}, {Lunney}, {Maire}, {Mayor}, {Micela}, {Ordway},
  {Peacock}, {Phillips}, {Piotto}, {Pollacco}, {Queloz}, {Rice}, {Riverol},
  {Riverol}, {San Juan}, {Sasselov}, {Segransan}, {Sozzetti}, {Sosnowska},
  {Stobie}, {Szentgyorgyi}, {Vick}, \& {Weber}}]{cos12}
{Cosentino}, R., {Lovis}, C., {Pepe}, F., {et~al.} 2012, in Society of
  Photo-Optical Instrumentation Engineers (SPIE) Conference Series, Vol. 8446,
  Society of Photo-Optical Instrumentation Engineers (SPIE) Conference Series,
  1

\bibitem[{{Cutri} {et~al.}(2003){Cutri}, {Skrutskie}, {van Dyk}, {Beichman},
  {Carpenter}, {Chester}, {Cambresy}, {Evans}, {Fowler}, {Gizis}, {Howard},
  {Huchra}, {Jarrett}, {Kopan}, {Kirkpatrick}, {Light}, {Marsh}, {McCallon},
  {Schneider}, {Stiening}, {Sykes}, {Weinberg}, {Wheaton}, {Wheelock}, \&
  {Zacarias}}]{cur03}
{Cutri}, R.~M., {Skrutskie}, M.~F., {van Dyk}, S., {et~al.} 2003, VizieR Online
  Data Catalog, 2246, 0

\bibitem[{{Danby}(1988)}]{dan88}
{Danby}, J.~M.~A. 1988, {Fundamentals of celestial mechanics} (2nd ed.;
  Richmond, Va., U.S.A.~: Willmann-Bell)

\bibitem[{{Davies} {et~al.}(2015){Davies}, {Bedding}, \& {Silva
  Aguirre}}]{dav15}
{Davies}, G.~R., {Bedding}, T., \& {Silva Aguirre}, V. e.~a. 2015, MNRAS,
  submitted

\bibitem[{{Davies} {et~al.}(2014){Davies}, {Handberg}, {Miglio}, {Campante},
  {Chaplin}, \& {Elsworth}}]{dav14}
{Davies}, G.~R., {Handberg}, R., {Miglio}, A., {et~al.} 2014, \mnras, 445, L94

\bibitem[{{Delfosse} {et~al.}(2000){Delfosse}, {Forveille}, {S{\'e}gransan},
  {Beuzit}, {Udry}, {Perrier}, \& {Mayor}}]{del00}
{Delfosse}, X., {Forveille}, T., {S{\'e}gransan}, D., {et~al.} 2000, \aap, 364,
  217

\bibitem[{{Demarque} {et~al.}(2008){Demarque}, {Guenther}, {Li}, {Mazumdar}, \&
  {Straka}}]{dem08}
{Demarque}, P., {Guenther}, D.~B., {Li}, L.~H., {Mazumdar}, A., \& {Straka},
  C.~W. 2008, \apss, 316, 31

\bibitem[{{Dressing} \& {Charbonneau}(2015)}]{dre15}
{Dressing}, C.~D., \& {Charbonneau}, D. 2015, \apj, 807, 45

\bibitem[{{Dressing} {et~al.}(2015){Dressing}, {Charbonneau}, {Dumusque},
  {Gettel}, {Pepe}, {Collier Cameron}, {Latham}, {Molinari}, {Udry}, {Affer},
  {Bonomo}, {Buchhave}, {Cosentino}, {Figueira}, {Fiorenzano}, {Harutyunyan},
  {Haywood}, {Johnson}, {Lopez-Morales}, {Lovis}, {Malavolta}, {Mayor},
  {Micela}, {Motalebi}, {Nascimbeni}, {Phillips}, {Piotto}, {Pollacco},
  {Queloz}, {Rice}, {Sasselov}, {S{\'e}gransan}, {Sozzetti}, {Szentgyorgyi}, \&
  {Watson}}]{dre15b}
{Dressing}, C.~D., {Charbonneau}, D., {Dumusque}, X., {et~al.} 2015, \apj, 800,
  135

\bibitem[{{Dumusque} {et~al.}(2014){Dumusque}, {Bonomo}, {Haywood},
  {Malavolta}, {S{\'e}gransan}, {Buchhave}, {Collier Cameron}, {Latham},
  {Molinari}, {Pepe}, {Udry}, {Charbonneau}, {Cosentino}, {Dressing},
  {Figueira}, {Fiorenzano}, {Gettel}, {Harutyunyan}, {Horne}, {Lopez-Morales},
  {Lovis}, {Mayor}, {Micela}, {Motalebi}, {Nascimbeni}, {Phillips}, {Piotto},
  {Pollacco}, {Queloz}, {Rice}, {Sasselov}, {Sozzetti}, {Szentgyorgyi}, \&
  {Watson}}]{dum14}
{Dumusque}, X., {Bonomo}, A.~S., {Haywood}, R.~D., {et~al.} 2014, \apj, 789,
  154

\bibitem[{{Els} {et~al.}(2014){Els}, {Lock}, {Comoretto}, {Gracia},
  {O'Mullane}, {Cheek}, \& {Vallenari}}]{els14}
{Els}, S., {Lock}, T., {Comoretto}, G., {et~al.} 2014, in Society of
  Photo-Optical Instrumentation Engineers (SPIE) Conference Series, Vol. 9150,
  Society of Photo-Optical Instrumentation Engineers (SPIE) Conference Series,
  0

\bibitem[{{Ford}(2005)}]{for05}
{Ford}, E.~B. 2005, \aj, 129, 1706

\bibitem[{{Ford}(2006)}]{for06}
---. 2006, \apj, 642, 505

\bibitem[{{Foreman-Mackey} {et~al.}(2013){Foreman-Mackey}, {Hogg}, {Lang}, \&
  {Goodman}}]{for13}
{Foreman-Mackey}, D., {Hogg}, D.~W., {Lang}, D., \& {Goodman}, J. 2013, \pasp,
  125, 306

\bibitem[{{Fortney} {et~al.}(2007){Fortney}, {Marley}, \& {Barnes}}]{for07}
{Fortney}, J.~J., {Marley}, M.~S., \& {Barnes}, J.~W. 2007, \apj, 659, 1661

\bibitem[{{Fressin} {et~al.}(2013){Fressin}, {Torres}, {Charbonneau}, {Bryson},
  {Christiansen}, {Dressing}, {Jenkins}, {Walkowicz}, \& {Batalha}}]{fre13}
{Fressin}, F., {Torres}, G., {Charbonneau}, D., {et~al.} 2013, \apj, 766, 81

\bibitem[{{Furesz}(2008)}]{fur08}
{Furesz}, G. 2008, PhD Thesis, Univ. of Szeged

\bibitem[{{Gelman} \& {Rubin}(1992)}]{gel92}
{Gelman}, A., \& {Rubin}, D.~B. 1992, Stat. Sci., 7, 457

\bibitem[{{Gilks} {et~al.}(1995){Gilks}, {Richardson}, \&
  {Spiegelhalter}}]{gil95}
{Gilks}, W.~R., {Richardson}, S., \& {Spiegelhalter}, D.~J. 1995, Markov Chain
  Monte Carlo in Practice (Boca Raton: Chapman \& Hall/CRC)

\bibitem[{{Gilliland} {et~al.}(2011){Gilliland}, {McCullough}, {Nelan},
  {Brown}, {Charbonneau}, {Nutzman}, {Christensen-Dalsgaard}, \&
  {Kjeldsen}}]{gil11}
{Gilliland}, R.~L., {McCullough}, P.~R., {Nelan}, E.~P., {et~al.} 2011, \apj,
  726, 2

\bibitem[{{Grasset} {et~al.}(2009){Grasset}, {Schneider}, \& {Sotin}}]{gra09}
{Grasset}, O., {Schneider}, J., \& {Sotin}, C. 2009, \apj, 693, 722

\bibitem[{{Handberg} \& {Lund}(2014)}]{han14}
{Handberg}, R., \& {Lund}, M.~N. 2014, \mnras, 445, 2698

\bibitem[{{Henning} {et~al.}(2009){Henning}, {O'Connell}, \&
  {Sasselov}}]{hen09}
{Henning}, W.~G., {O'Connell}, R.~J., \& {Sasselov}, D.~D. 2009, \apj, 707,
  1000

\bibitem[{{H{\o}g} {et~al.}(2000){H{\o}g}, {Fabricius}, {Makarov}, {Urban},
  {Corbin}, {Wycoff}, {Bastian}, {Schwekendiek}, \& {Wicenec}}]{hog00}
{H{\o}g}, E., {Fabricius}, C., {Makarov}, V.~V., {et~al.} 2000, \aap, 355, L27

\bibitem[{{Howard} {et~al.}(2009){Howard}, {Johnson}, {Marcy}, {Fischer},
  {Wright}, {Henry}, {Giguere}, {Isaacson}, {Valenti}, {Anderson}, \&
  {Piskunov}}]{how09}
{Howard}, A.~W., {Johnson}, J.~A., {Marcy}, G.~W., {et~al.} 2009, \apj, 696, 75

\bibitem[{{Huber} {et~al.}(2013){Huber}, {Chaplin}, {Christensen-Dalsgaard},
  {Gilliland}, {Kjeldsen}, {Buchhave}, {Fischer}, {Lissauer}, {Rowe},
  {Sanchis-Ojeda}, {Basu}, {Handberg}, {Hekker}, {Howard}, {Isaacson},
  {Karoff}, {Latham}, {Lund}, {Lundkvist}, {Marcy}, {Miglio}, {Silva Aguirre},
  {Stello}, {Arentoft}, {Barclay}, {Bedding}, {Burke}, {Christiansen},
  {Elsworth}, {Haas}, {Kawaler}, {Metcalfe}, {Mullally}, \& {Thompson}}]{hub13}
{Huber}, D., {Chaplin}, W.~J., {Christensen-Dalsgaard}, J., {et~al.} 2013,
  \apj, 767, 127

\bibitem[{{Kass} \& {Raftery}(1995)}]{kas95}
{Kass}, R.~E., \& {Raftery}, A.~E. 1995, Journal of the American Statistical
  Association, 430, 773

\bibitem[{{Kipping}(2013)}]{kipping2013}
{Kipping}, D.~M. 2013, \mnras, 435, 2152

\bibitem[{{Kipping}(2014)}]{kip14}
---. 2014, \mnras, 440, 2164

\bibitem[{{Kov{\'a}cs} {et~al.}(2002){Kov{\'a}cs}, {Zucker}, \&
  {Mazeh}}]{kov02}
{Kov{\'a}cs}, G., {Zucker}, S., \& {Mazeh}, T. 2002, \aap, 391, 369

\bibitem[{{L{\'e}ger} {et~al.}(2004){L{\'e}ger}, {Selsis}, {Sotin}, {Guillot},
  {Despois}, {Mawet}, {Ollivier}, {Lab{\`e}que}, {Valette}, {Brachet},
  {Chazelas}, \& {Lammer}}]{leg04}
{L{\'e}ger}, A., {Selsis}, F., {Sotin}, C., {et~al.} 2004, Icarus, 169, 499

\bibitem[{{Lopez} \& {Fortney}(2014)}]{lop14}
{Lopez}, E.~D., \& {Fortney}, J.~J. 2014, \apj, 792, 1

\bibitem[{{Lopez} {et~al.}(2012){Lopez}, {Fortney}, \& {Miller}}]{lop12}
{Lopez}, E.~D., {Fortney}, J.~J., \& {Miller}, N. 2012, \apj, 761, 59

\bibitem[{{Mamajek} \& {Hillenbrand}(2008)}]{mam08}
{Mamajek}, E.~E., \& {Hillenbrand}, L.~A. 2008, \apj, 687, 1264

\bibitem[{{Mandel} \& {Agol}(2002)}]{man02}
{Mandel}, K., \& {Agol}, E. 2002, \apjl, 580, L171

\bibitem[{{Marcus} {et~al.}(2010){Marcus}, {Sasselov}, {Hernquist}, \&
  {Stewart}}]{mar10}
{Marcus}, R.~A., {Sasselov}, D., {Hernquist}, L., \& {Stewart}, S.~T. 2010,
  \apjl, 712, L73

\bibitem[{{Marcy} {et~al.}(2014){Marcy}, {Isaacson}, {Howard}, {Rowe},
  {Jenkins}, {Bryson}, {Latham}, {Howell}, {Gautier}, {Batalha}, {Rogers},
  {Ciardi}, {Fischer}, {Gilliland}, {Kjeldsen}, {Christensen-Dalsgaard},
  {Huber}, {Chaplin}, {Basu}, {Buchhave}, {Quinn}, {Borucki}, {Koch}, {Hunter},
  {Caldwell}, {Van Cleve}, {Kolbl}, {Weiss}, {Petigura}, {Seager}, {Morton},
  {Johnson}, {Ballard}, {Burke}, {Cochran}, {Endl}, {MacQueen}, {Everett},
  {Lissauer}, {Ford}, {Torres}, {Fressin}, {Brown}, {Steffen}, {Charbonneau},
  {Basri}, {Sasselov}, {Winn}, {Sanchis-Ojeda}, {Christiansen}, {Adams},
  {Henze}, {Dupree}, {Fabrycky}, {Fortney}, {Tarter}, {Holman}, {Tenenbaum},
  {Shporer}, {Lucas}, {Welsh}, {Orosz}, {Bedding}, {Campante}, {Davies},
  {Elsworth}, {Handberg}, {Hekker}, {Karoff}, {Kawaler}, {Lund}, {Lundkvist},
  {Metcalfe}, {Miglio}, {Silva Aguirre}, {Stello}, {White}, {Boss}, {Devore},
  {Gould}, {Prsa}, {Agol}, {Barclay}, {Coughlin}, {Brugamyer}, {Mullally},
  {Quintana}, {Still}, {Thompson}, {Morrison}, {Twicken}, {D{\'e}sert},
  {Carter}, {Crepp}, {H{\'e}brard}, {Santerne}, {Moutou}, {Sobeck}, {Hudgins},
  {Haas}, {Robertson}, {Lillo-Box}, \& {Barrado}}]{mar14}
{Marcy}, G.~W., {Isaacson}, H., {Howard}, A.~W., {et~al.} 2014, \apjs, 210, 20

\bibitem[{{Mayor} \& {Udry}(2008)}]{may08}
{Mayor}, M., \& {Udry}, S. 2008, Physica Scripta Volume T, 130, 014010

\bibitem[{{Mayor} {et~al.}(2003){Mayor}, {Pepe}, {Queloz}, {Bouchy},
  {Rupprecht}, {Lo Curto}, {Avila}, {Benz}, {Bertaux}, {Bonfils}, {Dall},
  {Dekker}, {Delabre}, {Eckert}, {Fleury}, {Gilliotte}, {Gojak}, {Guzman},
  {Kohler}, {Lizon}, {Longinotti}, {Lovis}, {Megevand}, {Pasquini}, {Reyes},
  {Sivan}, {Sosnowska}, {Soto}, {Udry}, {van Kesteren}, {Weber}, \&
  {Weilenmann}}]{may03}
{Mayor}, M., {Pepe}, F., {Queloz}, D., {et~al.} 2003, The Messenger, 114, 20

\bibitem[{{McQuillan} {et~al.}(2013){McQuillan}, {Mazeh}, \&
  {Aigrain}}]{mcq13b}
{McQuillan}, A., {Mazeh}, T., \& {Aigrain}, S. 2013, \apjl, 775, L11

\bibitem[{{Meadows} \& {Seager}(2010)}]{meadows+seager2010}
{Meadows}, V., \& {Seager}, S. 2010, {Terrestrial Planet Atmospheres and
  Biosignatures}, ed. S.~{Seager}, 441--470

\bibitem[{Metropolis {et~al.}(1953)Metropolis, Rosenbluth, Rosenbluth, Teller,
  \& Teller}]{met53}
Metropolis, N., Rosenbluth, A.~W., Rosenbluth, M.~N., Teller, A.~H., \& Teller,
  E. 1953, The Journal of Chemical Physics, 21, 1087

\bibitem[{{Morton} \& {Johnson}(2011)}]{mor11}
{Morton}, T.~D., \& {Johnson}, J.~A. 2011, \apj, 738, 170

\bibitem[{{Motalebi} {et~al.}(2015){Motalebi}, {Udry}, {Gillon}, {Lovis},
  {Segransan}, {Buchhave}, {Demory}, {Malavolta}, {Dressing}, {Sasselov},
  {Rice}, {Charbonneau}, {Collier Cameron}, {Latham}, {Molinari}, {Pepe},
  {Affer}, {Bonomo}, {Cosentino}, {Dumusque}, {Figueira}, {Fiorenzano},
  {Gettel}, {Harutyunyan}, {Haywood}, {Johnson}, {Lopez}, {Lopez-Morales},
  {Mayor}, {Micela}, {Mortier}, {Nascimbeni}, {Philips}, {Piotto}, {Pollacco},
  {Queloz}, {Sozzetti}, {Vanderburg}, \& {Watson}}]{mot15}
{Motalebi}, F., {Udry}, S., {Gillon}, M., {et~al.} 2015, ArXiv e-prints,
  arXiv:1507.08532

\bibitem[{{Owen} \& {Jackson}(2012)}]{owe12}
{Owen}, J.~E., \& {Jackson}, A.~P. 2012, \mnras, 425, 2931

\bibitem[{{Pepe} {et~al.}(2002){Pepe}, {Mayor}, {Rupprecht}, {Avila},
  {Ballester}, {Beckers}, {Benz}, {Bertaux}, {Bouchy}, {Buzzoni}, {Cavadore},
  {Deiries}, {Dekker}, {Delabre}, {D'Odorico}, {Eckert}, {Fischer}, {Fleury},
  {George}, {Gilliotte}, {Gojak}, {Guzman}, {Koch}, {Kohler}, {Kotzlowski},
  {Lacroix}, {Le Merrer}, {Lizon}, {Lo Curto}, {Longinotti}, {Megevand},
  {Pasquini}, {Petitpas}, {Pichard}, {Queloz}, {Reyes}, {Richaud}, {Sivan},
  {Sosnowska}, {Soto}, {Udry}, {Ureta}, {van Kesteren}, {Weber}, {Weilenmann},
  {Wicenec}, {Wieland}, {Christensen-Dalsgaard}, {Dravins}, {Hatzes},
  {K{\"u}rster}, {Paresce}, \& {Penny}}]{pep02}
{Pepe}, F., {Mayor}, M., {Rupprecht}, G., {et~al.} 2002, The Messenger, 110, 9

\bibitem[{{Rowe} {et~al.}(2015){Rowe}, {Coughlin}, {Antoci}, {Barclay},
  {Batalha}, {Borucki}, {Burke}, {Bryson}, {Caldwell}, {Campbell},
  {Catanzarite}, {Christiansen}, {Cochran}, {Gilliland}, {Girouard}, {Haas},
  {He{\l}miniak}, {Henze}, {Hoffman}, {Howell}, {Huber}, {Hunter},
  {Jang-Condell}, {Jenkins}, {Klaus}, {Latham}, {Li}, {Lissauer}, {McCauliff},
  {Morris}, {Mullally}, {Ofir}, {Quarles}, {Quintana}, {Sabale}, {Seader},
  {Shporer}, {Smith}, {Steffen}, {Still}, {Tenenbaum}, {Thompson}, {Twicken},
  {Van Laerhoven}, {Wolfgang}, \& {Zamudio}}]{row15}
{Rowe}, J.~F., {Coughlin}, J.~L., {Antoci}, V., {et~al.} 2015, \apjs, 217, 16

\bibitem[{{Seager} {et~al.}(2007){Seager}, {Kuchner}, {Hier-Majumder}, \&
  {Militzer}}]{sea07}
{Seager}, S., {Kuchner}, M., {Hier-Majumder}, C.~A., \& {Militzer}, B. 2007,
  \apj, 669, 1279

\bibitem[{{Silva Aguirre} {et~al.}(2015){Silva Aguirre}, {Davies}, {Basu},
  {Christensen-Dalsgaard}, {Creevey}, {Metcalfe}, {Bedding}, {Casagrande},
  {Handberg}, {Lund}, {Nissen}, {Chaplin}, {Huber}, {Serenelli}, {Stello}, {Van
  Eylen}, {Campante}, {Elsworth}, {Gilliland}, {Hekker}, {Karoff}, {Kawaler},
  {Kjeldsen}, \& {Lundkvist}}]{sil15}
{Silva Aguirre}, V., {Davies}, G.~R., {Basu}, S., {et~al.} 2015, \mnras, 452,
  2127

\bibitem[{{Sliski} \& {Kipping}(2014)}]{sliski+kipping2014}
{Sliski}, D.~H., \& {Kipping}, D.~M. 2014, \apj, 788, 148

\bibitem[{{Sullivan} {et~al.}(2015){Sullivan}, {Winn}, {Berta-Thompson},
  {Charbonneau}, {Deming}, {Dressing}, {Latham}, {Levine}, {McCullough},
  {Morton}, {Ricker}, {Vanderspek}, \& {Woods}}]{sul15}
{Sullivan}, P.~W., {Winn}, J.~N., {Berta-Thompson}, Z.~K., {et~al.} 2015, \apj,
  809, 77

\bibitem[{{Tull} {et~al.}(1995){Tull}, {MacQueen}, {Sneden}, \&
  {Lambert}}]{tul95}
{Tull}, R.~G., {MacQueen}, P.~J., {Sneden}, C., \& {Lambert}, D.~L. 1995,
  \pasp, 107, 251

\bibitem[{{Tuomi} \& {Jones}(2012)}]{tuo12}
{Tuomi}, M., \& {Jones}, H.~R.~A. 2012, \aap, 544, A116

\bibitem[{{Valencia} {et~al.}(2006){Valencia}, {O'Connell}, \&
  {Sasselov}}]{val06}
{Valencia}, D., {O'Connell}, R.~J., \& {Sasselov}, D. 2006, Icarus, 181, 545

\bibitem[{{Vogt} {et~al.}(1994){Vogt}, {Allen}, {Bigelow}, {Bresee}, {Brown},
  {Cantrall}, {Conrad}, {Couture}, {Delaney}, {Epps}, {Hilyard}, {Hilyard},
  {Horn}, {Jern}, {Kanto}, {Keane}, {Kibrick}, {Lewis}, {Osborne},
  {Pardeilhan}, {Pfister}, {Ricketts}, {Robinson}, {Stover}, {Tucker}, {Ward},
  \& {Wei}}]{vog94}
{Vogt}, S.~S., {Allen}, S.~L., {Bigelow}, B.~C., {et~al.} 1994, in Society of
  Photo-Optical Instrumentation Engineers (SPIE) Conference Series, Vol. 2198,
  Instrumentation in Astronomy VIII, ed. D.~L. {Crawford} \& E.~R. {Craine},
  362

\bibitem[{{Weiss} \& {Schlattl}(2008)}]{wei08}
{Weiss}, A., \& {Schlattl}, H. 2008, \apss, 316, 99

\bibitem[{{White} {et~al.}(2011){White}, {Bedding}, {Stello},
  {Christensen-Dalsgaard}, {Huber}, \& {Kjeldsen}}]{whi11}
{White}, T.~R., {Bedding}, T.~R., {Stello}, D., {et~al.} 2011, \apj, 743, 161

\bibitem[{{White} {et~al.}(2012){White}, {Bedding}, {Gruberbauer}, {Benomar},
  {Stello}, {Appourchaux}, {Chaplin}, {Christensen-Dalsgaard}, {Elsworth},
  {Garc{\'{\i}}a}, {Hekker}, {Huber}, {Kjeldsen}, {Mosser}, {Kinemuchi},
  {Mullally}, \& {Still}}]{whi12}
{White}, T.~R., {Bedding}, T.~R., {Gruberbauer}, M., {et~al.} 2012, \apjl, 751,
  L36

\bibitem[{{Zeng} \& {Sasselov}(2013)}]{zen13}
{Zeng}, L., \& {Sasselov}, D. 2013, \pasp, 125, 227

\bibitem[{{Zhang} \& {Hamilton}(2008)}]{zha08}
{Zhang}, K., \& {Hamilton}, D.~P. 2008, \icarus, 193, 267

\end{thebibliography}

\begin{deluxetable}{lccc}
\tablecolumns{4}
\tabletypesize{}
\tablewidth{0pt}
\tablecaption{Stellar Parameters of Kepler-454\label{table:stellar}}
\setlength{\tabcolsep}{0.05in}
\tablehead{\colhead{Parameter} & \colhead{Value \& 1$\sigma$ Errors} & \colhead{Ref} }
\startdata
Right ascension & 19$^{h}$ 09$^{m}$ 54\arcs841  & \citet{hog00}\\
Declination &+38$^{d}$ 13$^{m}$ 43\arcs95  & \citet{hog00}\\	
Kepler magnitude & 11.457 & \citet{bor11}\\
$V$ magnitude & 11.57 & \citet{hog00} \\
$K_{s}$ magnitude & 9.968 & \citet{cur03}\\
log $g$ & 4.395 $^{+0.077}_{-0.055}$ & this work \\
$R_{*}$ ($R_{\odot}$) &1.066 $\pm$ 0.012 & this work \\
$M_{*}$ ($M_{\odot}$) &1.028 $^{+0.04}_{-0.03}$ & this work\\
$\rho_{*}$ (g cm$^{-3}$) & 1.199 $^{+0.015}_{-0.014}$& this work\\
Age (Gyr) &5.25$^{+1.41}_{-1.39}$& this work\\
\teff\ (K) & 5687 $\pm$ 49 & this work, SPC\\
$[$m/H$]$ & 0.32 $\pm$ 0.08 & this work, SPC \\
$v$ sin $i$  (km s$^{-1}$) & $<$2.4 & this work, SPC\\
log $g$ & 4.37 $\pm$ 0.06 & this work, MOOG \\
\teff\ (K) & 5701 $\pm$ 34 & this work, MOOG\\
$[$Fe/H$]$ & 0.27 $\pm$ 0.04 & this work, MOOG \\
\vmicro\ (km s$^{-1}$) & 0.98 $\pm$ 0.07 & this work, MOOG\\

\enddata
\end{deluxetable}

\begin{deluxetable}{cc}
\tablecaption{Estimated oscillation frequencies of Kepler-454 ($\rm \mu Hz$)} 
\tablewidth{0pt}
\tablehead{
\colhead{$l=0$}& \colhead{$l=1$}}
\startdata
$2305.79 \pm 0.94$& $2362.12 \pm 0.90$\\
$2428.16 \pm 1.19$& $2487.73 \pm 0.75$\\
$2552.64 \pm 0.95$&    ...   \\
$2677.31 \pm 0.43$& $2736.76 \pm 0.39$\\
$2801.95 \pm 0.30$& $2861.59 \pm 0.26$\\
$2926.45 \pm 0.22$& $2986.41 \pm 0.31$\\
$3051.43 \pm 1.49$&    ...         
\enddata
\label{tab:tab1}
\end{deluxetable}

\begin{deluxetable}{ccc}
\tablecolumns{3}
\tablewidth{0pt}
\scriptsize
\tablecaption{Transit Parameters of Kepler-454}
\tablehead{
\colhead{Parameter} &
\colhead{Median} & 
\colhead{Error}
}
\startdata
$P$ (days) & 10.57375339 &  7.77e-06 \\
$T_{\rm 0} ({\rm BJD}_{\rm UTC})$ & 2455008.0675855 &  0.0007718 \\
$a/R_\star$ &  18.293    &   1.098\\
$R_p/R_\star$ & 0.02041  & 0.0011  \\
$b$ & 0.9288  & 0.0091 \\
$i$ (deg) & 87.090 & 0.203 \\
$q_1$ & 0.489 & 0.094 \\
$q_2$ & 0.517 & 0.330 \\
$u_1$ & 0.709 & 0.480 \\
$u_2$ & -0.023 & 0.453 \\
$R_p$ ($R_{\oplus}$) & 2.37 & 0.13 
\enddata
\label{table:tparam}
\end{deluxetable}

\begin{deluxetable}{lcccccc}
\tablecolumns{7}
\tabletypesize{}
\tablewidth{0pt}
\tablecaption{HARPS-N Radial Velocity Measurements of Kepler-454\label{table:rvdat}\tablenotemark{1}}

\setlength{\tabcolsep}{0.05in}
\tablehead{\colhead{BJD$_{\rm UTC}$} & \colhead{Radial Velocity} & \colhead{$\sigma_{\rm RV}$} & \colhead{Bisector Span} & \colhead{log $R^{'}_{\rm HK}$} & \colhead{$\sigma$log $R^{'}_{\rm HK}$}& \colhead{t$_{\rm exp}$} \\
\colhead{- 2400000}& \colhead{(m s$^{-1}$)}& \colhead{(m s$^{-1}$)}& \colhead{(m s$^{-1}$)}& \colhead{(dex)}& \colhead{(dex)}& \colhead{(s)}}
\startdata
56813.668879&	-71408.25&	1.42&	-29.41&	-5.03&	0.02&	1800\\
56814.513734&	-71408.84&	1.78&	-40.04&	-5.01&	0.03&	1800\\
56815.522009&	-71408.40&	1.55&	-26.38&	-5.03&	0.02&	1800\\
56816.591592&	-71408.43&	1.58&	-29.14&	-5.02&	0.03&	1800\\
56828.676554&	-71400.04&	1.75&	-25.51&	-5.05&	0.03&	1800\\
...&              ...&          ...&    ...&     ...&    ...&   ...\\
\enddata

\tablenotetext{1}{(This table is available in its entirety in machine-readable form.)}
\end{deluxetable}

\begin{deluxetable}{lccc}
\tablecolumns{4}
\tabletypesize{}
\tablewidth{0pt}
\tablecaption{HIRES Radial Velocity Measurements of Kepler-454\label{table:rvdat2}\tablenotemark{1}}
\setlength{\tabcolsep}{0.05in}
\tablehead{\colhead{BJD$_{\rm UTC}$} & \colhead{Radial Velocity} & \colhead{$\sigma_{\rm RV}$} \\\colhead{- 2400000} & \colhead{(m s$^{-1}$)} & \colhead{(m s$^{-1}$)} }
\startdata

55431.809347	&  52.75	& 1.24\\
55782.924716	& -113.43	& 1.31\\
55792.863793	& -106.50	& 1.85\\
55792.895597	& -110.48	& 1.40\\
55797.795127	& -99.82	& 1.26\\
...&              ...&            ...\\

\enddata
\tablenotetext{1}{(This table is available in its entirety in machine-readable form.)}
\end{deluxetable}

\begin{deluxetable}{lcccc}
\tablecolumns{5}
\tabletypesize{}
\tablewidth{0pt}
\tablecaption{RV Parameters of Kepler-454\label{table:rvparam}}
\setlength{\tabcolsep}{0.05in}
\tablehead{\colhead{Parameter} & \colhead{Best Fit} & \colhead{Best Fit} & \colhead{Eccentric Fit} & \colhead{Eccentric Fit}\\\colhead{} & \colhead{Median} & \colhead{1$\sigma$ Error} & \colhead{Median} & \colhead{1$\sigma$ Error} }
\startdata
\cutinhead{Kepler-454b}
$P$ (days) & 10.573753 & 7.5$\times 10^{-6}$ & 10.573753& 7.5$\times 10^{-6}$\\
$T_{\rm 0}$ (BJD$_{\rm UTC}$) & 2455008.06758 & 0.00076 & 2455008.06758 & 0.00076\\
$e$ & 0.0 & --- &0.23 &0.13 \\
$\omega\ (\mbox{deg})$ & --- & --- & 341.6 & 52.8 \\
$K$ (m s$^{-1}$) & 1.96 & 0.38& 2.16&0.43\\
$m_{p}$ ($M_{\oplus}$) & 6.84 & 1.40& 7.24&1.40\\
$a$ (AU) & 0.0954 & 0.0012& 0.0954& 0.0012\\
\cutinhead{Kepler-454c}
$P$ (days) & 523.90 & 0.70&523.89 &0.78\\
$T_{p}$ (BJD$_{\rm UTC}$) & 2454892 &  26 &2454897 & 27\\
$e$ & 0.0214 & 0.0077 & 0.0199 &0.0064\\
$\omega\ (\mbox{deg})$ & 337.4 & 17.4 & 341.3& 18.2\\
$K$ (m s$^{-1}$) & 110.44 & 0.96 &110.65 &0.99\\
$m_{p}$ sin $i$ ($M_{J}$) & 4.46 & 0.12& 4.47&0.12\\ 
$a$ sin $i$ (AU) & 1.286 & 0.0166 & 1.286 & 0.0166\\
\cutinhead{Kepler-454 System}
$\gamma$ (m s$^{-1}$ d$^{-1}$) & -71322.23 & 0.61& -71321.91&0.61\\ 
$dv/dt$ (m s$^{-1}$ d$^{-1}$) &0.0429& 0.0016&0.0431 &0.0016\\
HIRES offset (m s$^{-1}$ d$^{-1}$) & -71324.77&0.95& -71324.85&0.95\\
\cutinhead{Epoch of Fit (BJD$_{\rm UTC}$) = 2456847.8981528}

\enddata
\end{deluxetable}



\end{document}